\pdfoutput=1
\documentclass[aip,reprint,bibnotes,groupedaddress,letterpaper,fleqn,flushbottom,balancelastpage]{revtex4-1}

\usepackage{amsmath}
\usepackage{graphicx}
\usepackage{xcolor}

\usepackage[margin=1.7cm]{geometry}
\usepackage{setspace}

\usepackage[justification=raggedright,font=footnotesize,singlelinecheck=false,tableposition=top]{caption}

\usepackage[T1]{fontenc}
\usepackage{mathptmx}

\usepackage{mathtools, nccmath}

\usepackage[
bookmarks=true,
bookmarksnumbered=true,
hypertexnames=false,
]{hyperref}

\hypersetup{breaklinks=true,
colorlinks =true,
plainpages =true,
linktocpage=true,
linkcolor=blue,
citecolor=blue,
urlcolor =blue,
pdfborder={0 0 0},
pdfauthor={Jetty, Suman, Khaparde},
pdftitle ={Novel cases of diffraction of light from a grating},
pdfsubject  = {Research Article},
pdfkeywords = {Oblique grating diffraction},
pdfcreator  = {LaTeX with hyperref package},
pdfproducer = {dvipdf}
}

\usepackage{fancyhdr}

\begin{document}

\title{{\Large Novel cases of diffraction of light from a grating:\\Theory and experiment}\medskip}
\author{{\large Ninad R. Jetty and Akash Suman}}
\affiliation{UM-DAE Centre for Excellence in Basic Sciences, Mumbai, India\bigskip}
\author{{\large Rajesh B. Khaparde}}
\email{rajesh@hbcse.tifr.res.in}
\affiliation{\mbox{Homi Bhabha Centre for Science Education, Tata Institute of Fundamental Research, Mumbai, 400005, India}}
\date{July 02, 2012}

\begin{abstract}
\centering\begin{minipage}{\dimexpr\paperwidth-5.9cm}
  A popular pedagogical approach for introducing diffraction is to assume normal incidence of light on a single slit or a plane transmission grating. Interesting cases of diffraction from a grating at orientations other than normal incidence remain largely unexplored. In this article we report our study of these unexplored cases, which was taken up as an undergraduate student project. We define various cases of orientation of the grating, and use the Fresnel-Kirchhoff formula to arrive at the diffracted intensity distribution. An experimental arrangement consisting of a laser, a grating mount, a digital camera, and a calibrated plane screen is employed to record our observations. We discuss for each case the theoretical and experimental results, and establish the conformity between the two. Finally, we analyse the details of various cases and conclude that for an arbitrary orientation of the grating, the diffraction maxima fall along a second degree curve.\\
\noindent[\url{https://dx.doi.org/10.1119/1.4737854}]
\end{minipage}
\end{abstract}

\maketitle

\setstretch{0.97}
\vspace{-0.3cm}

\section{Introduction}
\vspace{-0.3cm}
The phenomenon of diffraction is ubiquitous when studying the wave nature of light. Diffraction for the case of normal incidence of light on a plane transmission grating has been well represented in the literature. When a laser beam is incident on a grating with vertical grooves, the diffraction maxima are seen as intense bright spots which fall along a horizontal line on a plane screen. The equation giving the condition for intensity maxima is derived from the intensity distribution, and is referred to as the grating equation.\cite{halliday,ghatak} As a slight modification of this case one may consider the grating to be tilted by an angle such that the grooves of the grating remain normal to the incident light. This breaks the symmetry of the diffraction pattern in the earlier case. Diffraction maxima on one side of the central maximum come close together while those on the other side go away from each other. Here, the condition for maxima is a modified form of the grating equation.\cite{feynman,michelson}

A study of the diffraction pattern when the grooves of the grating are inclined along the direction of the incident light has been carried out by Phadke and Allen.\cite{phadke} It is referred to as a case of ``oblique incidence.'' The diffraction maxima in this case fall along a curve and are symmetrically placed about the central maximum. This curve tends to become circular as the angle of inclination, as measured from the normal orientation of the grating, tends to $90^\textup{o}$. Phadke and Allen derive the locus of maxima and the intensity distribution on a flat screen, and give experimental results for this case.

In this article, we report our study of those novel cases of diffraction which arise when the grating is arbitrarily tilted, when the grooves may or may not remain normal to the incident light. Our objective is to carry out a theoretical analysis of these novel cases and to test the theoretical predictions against experimental observations. The coordinate systems we set up in Sec.~\ref{orientation} will help us define the different cases which can be rigorously analysed. In Sec.~\ref{theory}, we go through the steps that would be part of a typical derivation of the intensity distribution. In Sec.~\ref{expt} we describe the complete setup employed for experimental observations. Section~\ref{results} gives the theoretical and experimental results for different cases. In Sec.~\ref{discussion} we discuss the results and peculiar features of the diffraction pattern for each case, before concluding in Sec.~\ref{conclusion}.
\vspace{-0.4cm}

\section{Various Cases of Orientation} \label{orientation}
\begin{figure}[t]
\centering
\includegraphics[width=3.4in]{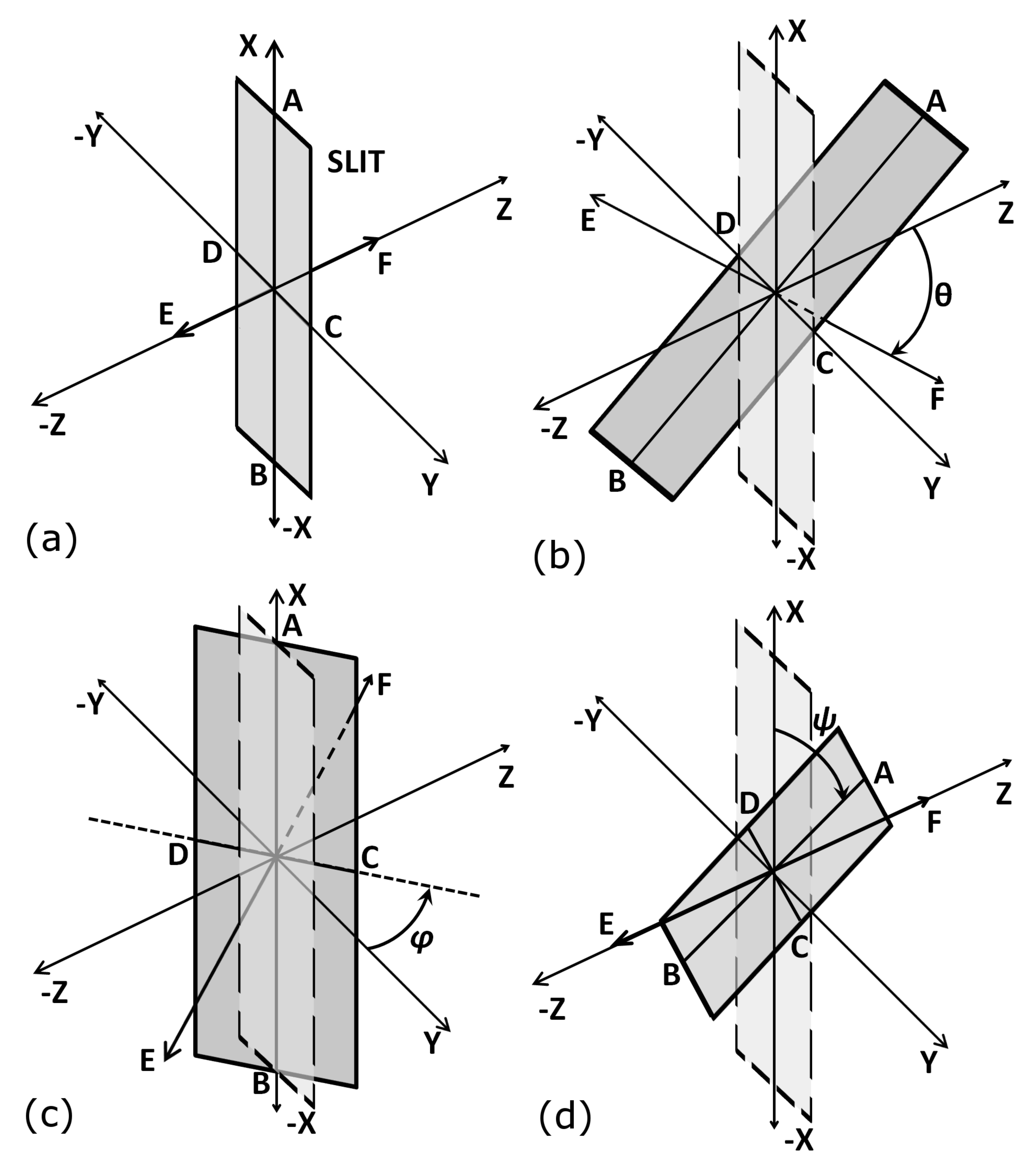}
\caption{(a) The coordinate system $X$-$Y$-$Z$ and the axis system $AB$-$CD$-$EF$ that is fixed to the slit. Tilting the slit about (b) axis $CD$ by angle $\theta$; (c) axis $AB$ by angle $\varphi$; and (d) axis $EF$ by angle $\psi$.}
\label{axesthetaphipsi}
\end{figure}

\vspace{-0.3cm}
We first carry out the analysis for a single slit\cite{rect} and then generalize the results for a grating of period $c$. Hence, the coordinate systems are defined for a single slit, but are also applicable to the grating. Consider a rectangular coordinate system $X$-$Y$-$Z$ as illustrated in Fig.~\hyperref[axesthetaphipsi]{\ref{axesthetaphipsi}(a)}. Using this system we can identify points on the surface of a slit in the grating. Points on neighboring slits can be identified by adding the grating period $c$ to the $y$~component of the position vector.

Another set of axes $AB$-$CD$-$EF$, which are fixed to the slit, is used to orient the slit in three-dimensional space. The slit can be tilted about any one of the axes $CD$, $AB$, or $EF$, the rotation being marked by the angles $\theta$, $\varphi$, or $\psi$ respectively as illustrated in Figs.~\hyperref[axesthetaphipsi]{\ref{axesthetaphipsi}(b)--\ref{axesthetaphipsi}(d)}. The arrows depict the sense of tilt marked by positive values (between $0^{\textup{o}}$ and $90^{\textup{o}}$) of the respective angle. Tilting the slit about $CD$ causes $EF$ and $AB$ to rotate in the $X$-$Z$ plane. Tilting the slit about $AB$ causes $CD$ and $EF$ to rotate in the $Y$-$Z$ plane. Tilting the slit about $EF$ causes $AB$ and $CD$ to rotate in the $X$-$Y$ plane.

The fact that the axis system $AB$-$CD$-$EF$ moves along with the slit as it tilts allows for tilting the slit successively by angles $\theta$, $\varphi$, or $\psi$ in various permutations. To uniquely fix the plane of an arbitrarily oriented slit in three-dimensional space it suffices to consider only two angles out of the three. We thus obtain six permutations of these angles, and this number cannot be further reduced because angular displacements do not commute. A general orientation of the grating can be obtained using any one of these six permutations.

Including the cases of normal incidence, tilting through angles $\theta$, $\varphi$, $\psi$, and their six permutations, we have a total of ten cases to discuss, which have been enumerated as elements of the set $S$ given below (the element $NI$ corresponds to the case of normal incidence):
\begin{align}
S &= \left \lbrace NI,\, \theta ,\, \varphi ,\, \psi ,\, \theta + \varphi ,\, \theta + \psi ,\, \varphi + \theta ,\right. \nonumber \\
&\quad \left.\, \varphi + \psi ,\, \psi + \theta ,\, \psi + \varphi \right \rbrace.
\end{align}
\vspace{-1.1cm}

\section{Common Theoretical Treatment} \label{theory}
\vspace{-0.3cm}

This section describes the theoretical treatment that is applicable to all the cases enumerated in set $S$. We use the Fresnel-Kirchhoff formula\cite{hecht,thyagarajan} which mathematically summarizes the Huygens-Fresnel Principle\cite{drude,brooker} of secondary wavelet construction. In the Fraunhofer diffraction limit, the Fresnel-Kirchhoff formula simplifies to an integral of the path difference of plane waves over the extent of the aperture.\cite{born,fowles} The final result is the net diffracted amplitude at a point on the screen.

\begin{figure}[!h]
\includegraphics[width=5.1in]{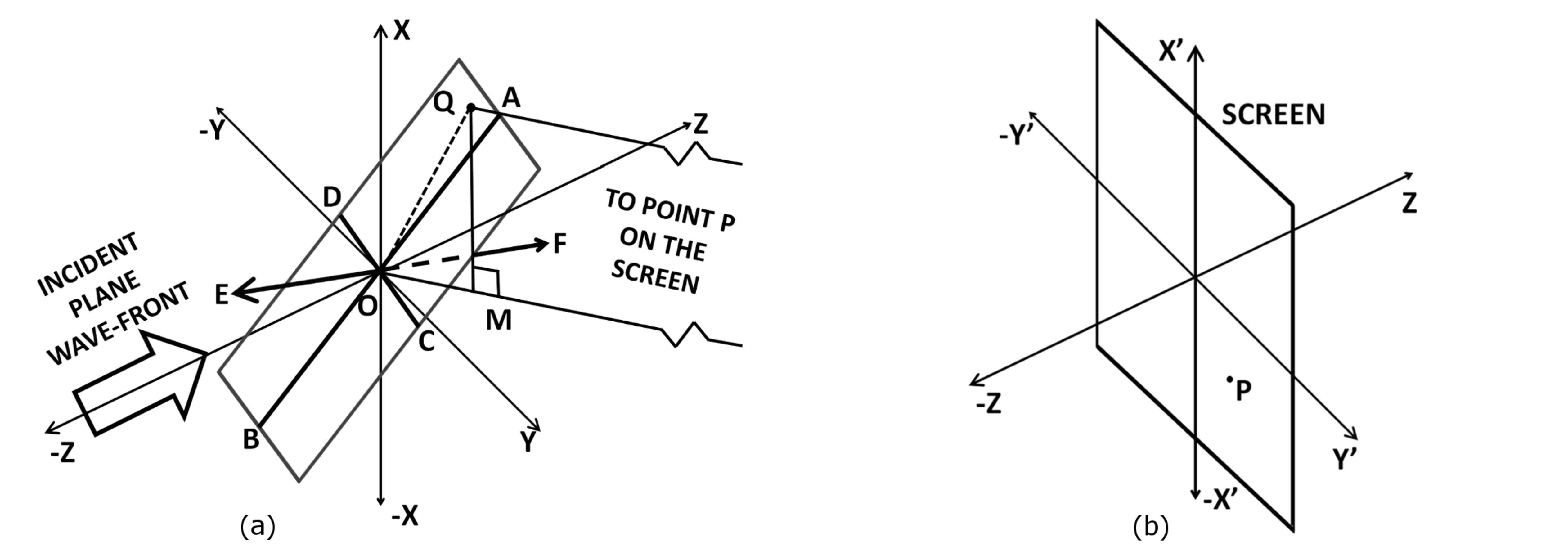}
\caption{(a) Schematic of the geometry used in the integral of the Fresnel-Kirchhoff formula. The direction of the incident light is from the $-Z$ to~$Z$. The infinitesimal area element around point $Q$ is given by $dw\,dl$ in the formula. Specific to every case, the coordinates of $Q$ will be functions of the two angles involved. (b) The $X^\prime$-$Y^\prime$ coordinate system used to identify points on the screen, which is placed at distance $D^{\prime}$ from the slit.}
\label{derivation}
\end{figure}

To obtain the amplitude distribution due to a single slit, consider a point $Q\equiv Q\left(Q_{x}\,,Q_{y}\,,Q_{z}\right)$ on the surface of the slit, as illustrated in Fig.~\ref{derivation}. Plane wave-fronts are incident along the $-Z$ to $Z$ direction. The path difference between the waves reaching point $P\equiv P\left(X^{\prime},Y^{\prime},D^{\prime}\right)$ on the screen from the origin $O$ and $Q$ has two contributions.\cite{phadke} One is the initial path difference between $O$ and $Q$, which is indicative of the delay between the incidence of a plane wave-front (parallel to the $X$-$Y$ plane) at the two points $O$ and $Q$, and the other is the path difference between rays $OP$ and $QP$. We identify these two contributions as the $Z$ coordinate of $Q$ (i.e., $Q_{z}$) and the projection of $OQ$ on $OP$ (i.e., $OM$ as in Fig.~\ref{derivation}) respectively. Note that $OM$ alone contributes to the path difference in the case of normal incidence, and that deviating from normal incidence introduces the additional path difference $Q_{z}$. The total path difference $\left(QP-OP\right)$ is then $\left(Q_{z}-OM\right)$ and the amplitude distribution $(U)$ is given by the Fresnel-Kirchhoff formula as\cite{phadke}
\begin{equation}
U =H \!\int_\textrm{area} \!\!\! dw \, dl \, \exp \left[ i k \left( Q_{z}-OM \right) \right], \label{amp}
\end{equation}
where $k$ is the wavenumber and $H$  is a constant. Performing this integral for any case of orientation and rearranging appropriately gives the amplitude distribution due to a single slit on the plane screen in the following form:\cite{longhurst,sirohi}
\begin{equation}
U = H^{\prime} \left( \frac{ \sin \beta_{l} }{\beta_{l}} \right) \left( \frac{\sin \beta_{w} }{\beta_{w}} \right). \label{rect}
\end{equation}
Here  $\beta_{l}$ and $\beta_{w}$ are, respectively, factors related to the length $(L)$ and width $(W)$ of the slit, which vary from case to case, while  $H^{\prime}=H\left(2L\right)\left(2W\right)$ for all the ten cases. The intensity distribution due to a single slit may be obtained by squaring Eq.~\eqref{rect}.

The amplitude distribution due to a grating with grating period $c$, of which $N$ slits out of the total have been illuminated, is given by the addition of the $N$ single-slit amplitude distributions.\cite{jenkins} The path differences involved are given in terms of the grating period, again with two distinct contributions. One is the path difference between any two adjacent slits at point $P$, and the other is indicative of the delay between incidence of a plane wave-front on two different slits.
It should be no surprise that these are similar to the two contributions we identified previously. Squaring this summed amplitude distribution gives the intensity distribution $(I)$ due to a grating in the form\cite{longhurst}
\begin{equation}
I = \left( H^{\prime} \right)^{2} \left( \frac{\sin \beta_{l} }{\beta_{l}} \right)^{2} \left( \frac{\sin \beta_{w}}{\beta_{w}} \right)^{2} \left( \frac{\sin \left( N \beta_{c}/2 \right)}{\sin \left( \beta_{c}/2 \right)} \right)^{2}, \label{grating}
\end{equation}
where $\beta_{c}$ is related to the grating period, and has the same form as $\beta_{w}$, with the slit width $W$ replaced by the grating period $c$. Note that the multiple slit (grating) intensity distribution can also be written as the product of the single slit intensity
\onecolumngrid
\,
\twocolumngrid
\noindent distribution and the ``interference term,'' $ \sin^{2} \left( N \beta_{c}/2 \right)/\sin^{2} \left( \beta_{c}/2 \right)$, of the $N$ slits.

Having identified the coordinates of the general point $Q$, one can use the treatment outlined above to arrive at the intensity distribution due to a grating for any case (marked by elements of set $S$) in the form of Eq.~\eqref{grating}.\cite{completederivations} The expressions for $\beta_{l}$, $\beta_{w}$, and $\beta_{c}$ are different for different cases. The power of the Fresnel-Kirchhoff formalism is manifest in reducing the problem of obtaining the intensity distribution to a mere geometric one, that of identifying the coordinates of $Q$.

Note how the Fresnel-Kirchhoff formula in the Fraunhofer diffraction limit suggests that the amplitude distribution is the Fourier transform of the aperture function.\cite{srivastava} It also generalizes the idea of diffraction of light, that it is the phenomenon that occurs when light encounters an inhomogeneity in its path. The inhomogeneity referred to above may be anything that alters the phase of the light wave, which in turn modifies the integral. It is essential to note that the formalism \textit{per se} does not assume a light wave and further reveals the universality of diffraction. Diffraction is thus not restricted to any particular kind of wave; it is interwoven with the wave nature itself. It is the association of light with an electromagnetic wave that gives the familiar observation of ``deviation of light from rectilinear propagation''\cite{hecht} made by Grimaldi.
\vspace{-0.3cm}

\section{Experimental Setup} \label{expt}
\vspace{-0.3cm}
The theory described in Sec.~\ref{orientation} requires the laser beam to be incident at the geometric center of the grating, which coincides with the origins of the axis systems $X$-$Y$-$Z$ and $AB$-$CD$-$EF$. Thus, the most important part of the experimental setup is the grating mount, employing a gyroscope-like arrangement of rings within a ring, essential for studying cases with a permutation of two angles. In such a permutation, the ring that allows tilting by the second angle must be within the ring that allows tilting by the first. For example, to obtain the orientation of $\theta + \varphi$, the $\varphi$~ring must be within the $\theta$~ring, while for the $\varphi + \theta$ orientation, the $\theta$~ring must be within the $\varphi$~ring. It is clear that there must be more than one ring for each angle, and hence a large number of inter-connected rings are required. We designed a grating mount as illustrated in Fig.~\hyperref[mountsetup]{\ref{mountsetup}(a)}, with the following five rings: outer $\psi$~ring, outer $\varphi$~ring, middle $\theta$~ring, inner $\varphi$~ring, and inner $\psi$~ring. This is the only configuration with the minimum number of rings that can allow the grating to be oriented as required for all the ten cases on the same mount.\cite{arrangement}

The grating mount consists of a wooden base that is attached to a wedge by hinges and two semicircular locking clamps, allowing the base to be tilted to obtain the outermost $\psi$~ring. To maintain the grating at the center of rings, during the experiment, the laser light is made to be incident on the grating after adjusting the outer $\psi$~ring. A circular disc that is mounted with its axle connected to the base constitutes the outer $\varphi$~ring. The middle $\theta$~ring is a semicircular metallic arc mounted on two rods fixed to the circular disc. The inner $\varphi$~ring is yet another metallic arc mounted from its center on the middle $\theta$~ring. A grating post with adjustable length holds the grating at the common center of the rings, as required. The innermost $\psi$~ring is obtained by sliding the base of this grating post over the inner $\varphi$~ring.

The complete experimental setup is illustrated in Fig.~\hyperref[mountsetup]{\ref{mountsetup}(b)}.  It consists of a green (532~nm) semiconductor laser source with a spot size of about 2~mm, a 12 lines per mm grating (ruling) mounted on the grating post of the grating mount, and a plane screen kept at a distance $D^{\prime}$ (about 1.3~m) from the grating. A digital camera is used to capture images of the diffraction pattern. A square grid (5~cm $\times$ 5~cm) is made on the screen to facilitate the overlay of observed patterns on the theoretically simulated ones.
\vspace{-0.4cm}

\section{Theoretical and Experimental Results} \label{results}
\vspace{-0.3cm}

Preliminary observations and simulations\cite{completederivations} brought out the difficulty of identifying a variable for observations, that is common to the diffraction pattern of all the ten cases, and yet uniquely identifies the pattern. Variables that are ideal for observations in some cases cannot be even appropriately defined for others, while certain other variables do not capture the essence of diffraction. For example, the distance between the 0th-order maximum of the transmitted and reflected beams on the screen is a feature that can be explained by reflection only; it is not peculiar to diffraction. Phadke and Allen have used this as an observation for the case of angle $\theta$. But they also derive the expression for the locus of maxima, which we figured we could use in our study.

The locus of diffraction maxima is given by the locus of points on the screen equidistant from the end points $A$ and $B$ of the slit\cite{phadke} [see the discussion after Eq.~\eqref{locus}]. The coordinates of $A$ and $B$ are derivable from those of $Q$ by substituting $w=0$ and $l=\pm L$ respectively [see Fig.~\hyperref[derivation]{\ref{derivation}(a)}]. For plane incident wave-fronts parallel to the $X$-$Y$ plane, point $A$ has an initial path difference with respect to $B$ given by twice the $Z$ coordinate of $A$. Including this initial path difference, the equation of locus is
\begin{equation}
\left( AP + 2Z \right) = BP \label{locus}.
\end{equation}
\onecolumngrid
\,
\begin{figure}[h!]
\includegraphics[width=5.3in]{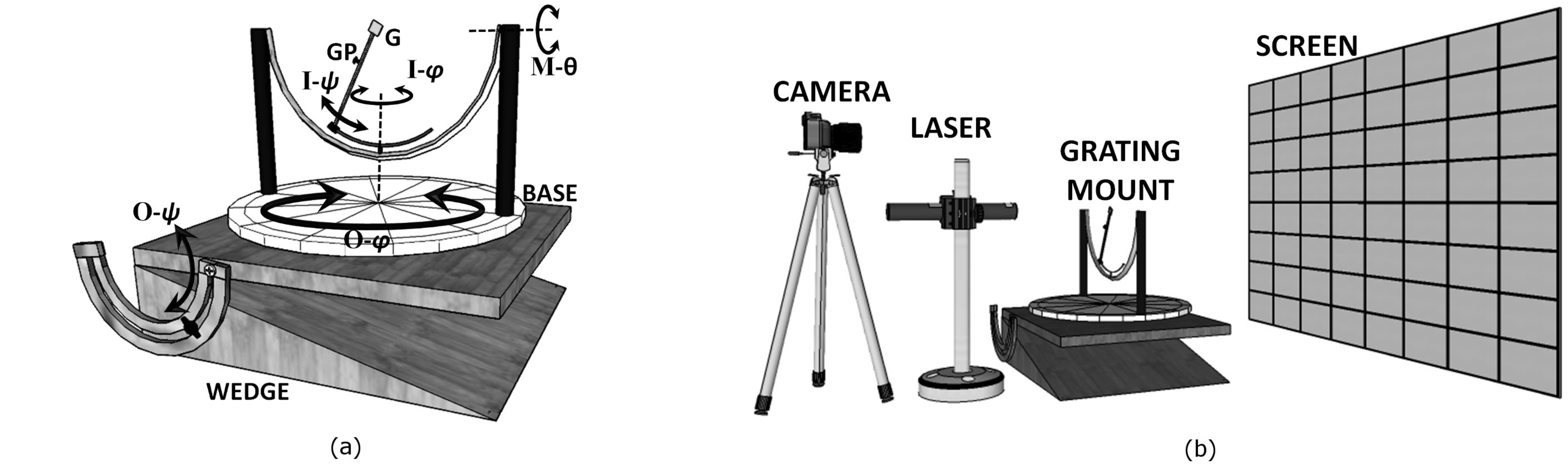}
\caption{(a) A 3-D schematic of the grating mount. G: grating, GP: variable-length grating post, O-$\psi$: outer $\psi$ ring, O-$\varphi$: outer $\varphi$ ring, M-$\theta$: middle $\theta$ ring, I-$\varphi$: inner $\varphi$ ring, I-$\psi$: inner $\psi$ ring. (b) A schematic of the complete experimental setup.}
\label{mountsetup}
\end{figure}
\twocolumngrid

When a slit is tilted, the coordinates of $A$ and $B$ change and hence the equation of locus is altered. This also leads to a change in the effective length and width of the slit as given by their respective projections on the $X$-$Y$ plane. Equation~\eqref{locus} implies that the rays from the ends of the slit interfere constructively with \textit{zero} phase difference, making the locus independent of the wavelength used. In a grating the slits have length much larger than the width, and hence other loci with nonzero $\left(2n\pi\right)$ phase difference are suppressed in intensity.

\begin{table}[b!]
\small
\caption{The values of angles $\theta$, $\varphi$, and $\psi$ used for observations and simulations. The screen is kept at a distance $D^{\prime}$ from the grating. The maximum uncertainties in the measurements of the angle and distance $D^{\prime}$ are about $0.5^{\textup{o}}$ and $1$~cm respectively.}
\label{data}
\begin{tabular}{lcccc}
\hline
\hline 
Case\hspace*{0.14in} & \hspace*{0.14in} $\theta \left(\deg\right)$\hspace*{0.14in} & \hspace*{0.14in} $\varphi \left(\deg\right)$\hspace*{0.14in} & \hspace*{0.14in} $\psi \left(\deg\right)$\hspace*{0.14in} & \hspace*{0.14in} $D^{\prime} \left(\right.$m$\left.\right)$ \\ 
\hline 
NI & 0.0 & 0.0 & 0.0 & 1.42 \\ 
$\theta$ & 81.1 & 0.0 & 0.0 & 1.43 \\ 
$\varphi$ & 0.0 & -60.0 & 0.0 & 1.42 \\ 
$\psi$ & 0.0 & 0.0 & -18.2 & 1.34 \\ 
$\theta + \varphi$ & 42.0 & 72.6 & 0.0 & 1.44 \\ 
$\theta + \psi$ & 82.4 & 0.0 & -15.0 & 1.44 \\ 
$\varphi + \theta$ & 82.4 & 15.0 & 0.0 & 1.42 \\ 
$\varphi + \psi$ & 0.0 & 75.0 & -37.0 & 1.20 \\ 
$\psi + \theta$ & 80.1 & 0.0 & -18.2 & 1.34 \\ 
$\psi + \varphi$ & 0.0 & 70.0 & -18.0 & 1.19 \\ 
\hline
\hline 
\end{tabular}
\vspace{+0.5cm}
\end{table}

We have obtained the expressions for the locus of maxima, for all the ten cases. Because the equation of locus [Eq.~\eqref{locus}] does not contain any term pertaining to the slit width, grating period, or number of slits illuminated, the locus is independent of all these factors. With a change in wavelength and the factors mentioned above, the separation between the maxima changes such that they move along the locus, which essentially remains the same. The locus, being determined only by the orientation of the grating and not other factors, is therefore a suitable choice for quantitative experimental observations.

With the experimental setup as described in Sec.~\ref{expt} (Fig.~\ref{mountsetup}), a photograph of the diffraction pattern on the screen is captured with the digital camera. As a test of theory, we check whether or not the observed pattern obeys the equation of locus [Eq.~\eqref{locus}]. Instead of solving the nonlinear equation for every case, we simulate the function $\left( AP + 2Z \right) - BP$  for all points $P$ on the screen, and plot a contour map\cite{completederivations} of the same using the software application \textsc{octave}.\cite{octave} In the contour map, lower values of the function are represented by darker contour lines, and higher values by lighter contour lines. We expect the experimentally observed pattern to fall along the minimum of this function, which is located between the two darkest contour lines. The contour lines do not have any physical significance themselves; they are a visual aid to indicate the minimum of the function, which is the physically significant locus of diffraction maxima.

The grids on the screen captured in the image act as a reference when overlaying the observed pattern and simulated contour plots using the software application \textsc{gimp}.\cite{gimp} These overlaid images\cite{completederivations} for every case test the agreement between theory and experiment. Note that this kind of observation is common to all ten cases.

In Subsections \hyperref[results-A]{\ref{results-A}}--\hyperref[results-J]{\ref{results-J}}, we consider the cases that we anticipated in Sec.~\ref{axesthetaphipsi}, enumerated as elements of the set~$S$. Table~\ref{data} gives the values of the angles that were used for the measurements and the simulations of the locus. As the general procedure for arriving at the intensity distributions has been outlined in detail in Sec.~\ref{orientation}, for each case it should suffice to give the expressions for $Q$, $\beta_{l}$, and $\beta_{w}$ and the equation of the locus in the form of Eq.~\eqref{locus}. The expressions for $\beta_{c}$ have not been specified as they are throughout similar to that of $\beta_{w}$, as mentioned in Sec.~\ref{orientation}.\cite{completederivations} The overlaid images of theoretical and observed locus are presented subsequent to the discussion of every two cases. In Figs.~\ref{nitheta}--\ref{psithetapsiphi}, the dark dots are experimentally recorded diffraction maxima and the continuous curves are the simulated contour lines.
\vspace{-0.3cm}

\subsection{Normal incidence}
\label{results-A}
\vspace{-0.5cm}
\begin{subequations}
\setlength{\jot}{0.2cm}
\begin{align}
&Q = Q \left[ l, w, 0 \right], \label{niq} \\
&\beta_{l} = kL \cos\left(\alpha_{x}\right), \label{nil} \\
&\beta_{w} = kW \cos\left(\alpha_{y}\right). \label{niw}
\end{align}
\end{subequations}
\vspace{-0.2cm}
The equation of locus is:
\begin{equation}
 \left\lbrace \left[X^{\prime}-L\right]^{2} + Y^{\prime2} + D^{\prime2} \right\rbrace^{1/2} 
= \left\lbrace \left[X^{\prime}+L\right]^{2} + Y^{\prime2} + D^{\prime2} \right\rbrace^{1/2}. \label{nie}
\end{equation}
\vspace{-1.5cm}

\subsection{Variation of only $\theta$}
\vspace{-0.5cm}
\begin{subequations}
\setlength{\jot}{0.2cm}
\begin{align}
&Q = Q \left[ l\cos\left(\theta\right), w, l\sin\left(\theta\right)\right], \label{tq} \\
&\beta_{l} = kL \left\lbrace \sin\left(\theta\right) \left[1-\cos\left(\alpha_{z}\right)\right] - \cos\left(\theta\right)\cos\left(\alpha_{x}\right) \right\rbrace, \label{tl} \\
&\beta_{w} = kW \cos\left(\alpha_{y}\right). \label{tw}
\end{align}
\end{subequations}
\vspace{-0.2cm}
The equation of locus is:
\begin{align}
&\left\lbrace \left[X^{\prime}-L\cos\left(\theta\right)\right]^{2} + Y^{\prime2} + \left[D^{\prime}-L \sin\left(\theta\right)\right]^{2} \right\rbrace^{1/2} + 2L\sin\left(\theta\right) \nonumber \\
&= \left\lbrace \left[X^{\prime}+L\cos\left(\theta\right)\right]^{2} + Y^{\prime2} + \left[D^{\prime}+L \sin\left(\theta\right)\right]^{2} \right\rbrace^{1/2}. \label{te}
\end{align}
\vspace{-0.8cm}

\subsection{Variation of only $\varphi$}
\vspace{-0.5cm}
\begin{subequations}
\setlength{\jot}{0.2cm}
\begin{align}
&Q = Q \left[ l, w\cos\left(\varphi\right), w\sin\left(\varphi\right) \right], \label{vq} \\
&\beta_{l} = kL \cos\left(\alpha_{x}\right),  \label{vl} \\
&\beta_{w} = kW \left\lbrace \sin\left(\varphi\right)\left[1-\cos\left(\alpha_{z}\right)\right] - \cos\left(\varphi\right)\cos\left(\alpha_{y}\right) \right\rbrace. \label{vw}
\end{align}
\end{subequations}
The equation of locus is same as that for the case of normal incidence, Eq.~\eqref{nie}.
\vspace{-0.3cm}

\subsection{Variation of only $\psi$}
\vspace{-0.5cm}
\begin{subequations}
\setlength{\jot}{0.2cm}
\begin{align}
&Q = Q \left[ l\cos\left(\psi\right)-w\sin\left(\psi\right), l\sin\left(\psi\right)+w\cos\left(\psi\right), 0 \right], \label{sq} \\
&\beta_{l} = kL \left\lbrace \cos\left(\psi\right)\cos\left(\alpha_{x}\right) + \sin\left(\psi\right)\cos\left(\alpha_{y}\right) \right\rbrace, \label{sl} \\
&\beta_{w} = kW \left\lbrace \sin\left(\psi\right)\cos\left(\alpha_{x}\right) - \cos\left(\psi\right)\cos\left(\alpha_{y}\right) \right\rbrace. \label{sw}
\end{align}
\end{subequations}
The equation of locus is:
\begin{align}
& \left\lbrace \left[X^{\prime}-L\cos\left(\psi\right)\right]^{2} + \left[Y^{\prime}-L\sin\left(\psi\right)\right]^{2} + D^{\prime2} \right\rbrace^{1/2} \nonumber \\
&= \left\lbrace \left[X^{\prime}+L\cos\left(\psi\right)\right]^{2} + \left[Y^{\prime}+L\sin\left(\psi\right)\right]^{2} + D^{\prime2} \right\rbrace^{1/2}. \label{se}
\end{align}
\newpage

\onecolumngrid

\begin{figure}[h!]
\includegraphics[width=4.8in]{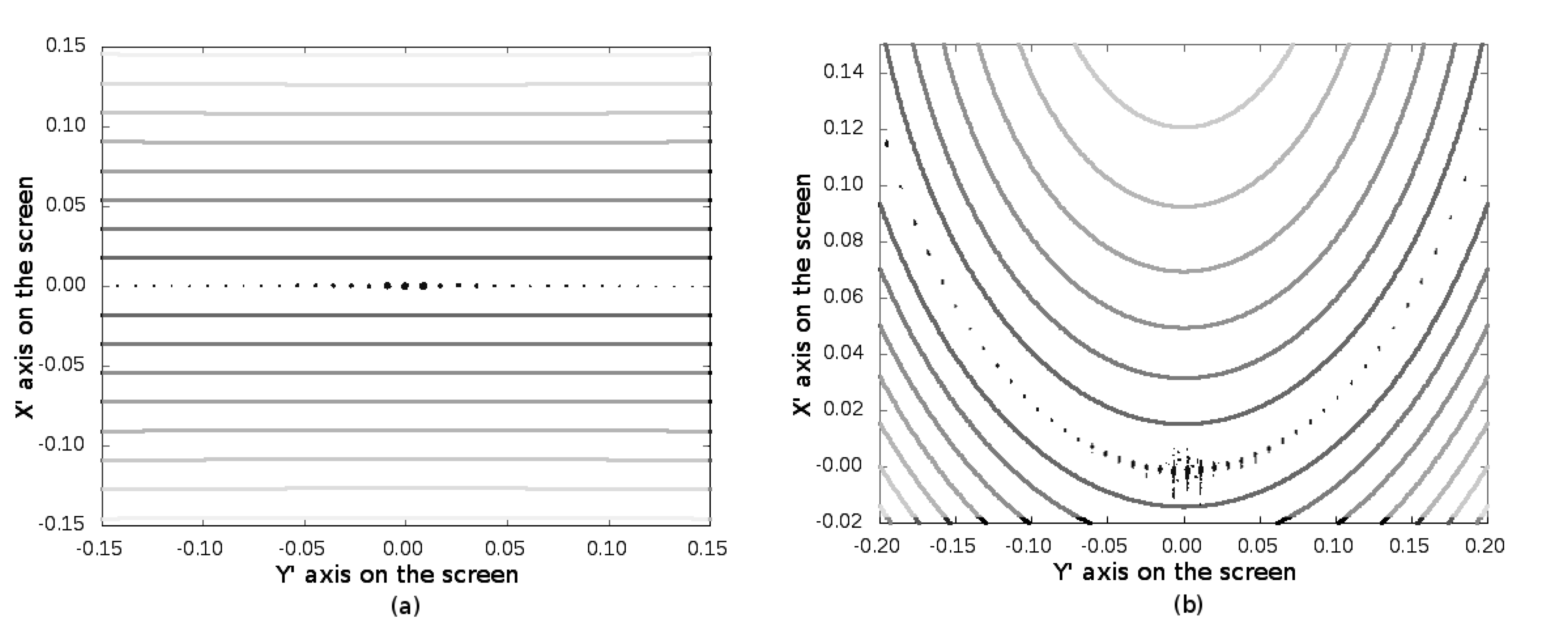}
\caption{Overlay of simulated and observed plots for (a) NI and (b) variation of only $\theta$. In the first case all three angles are zero; for the second case the value of $\theta$ is $81.1^{\textup{o}}$, the other two angles being zero.}
\label{nitheta}
\end{figure}
\vspace{-0.3cm}

\begin{figure}[h!]
\includegraphics[width=4.8in]{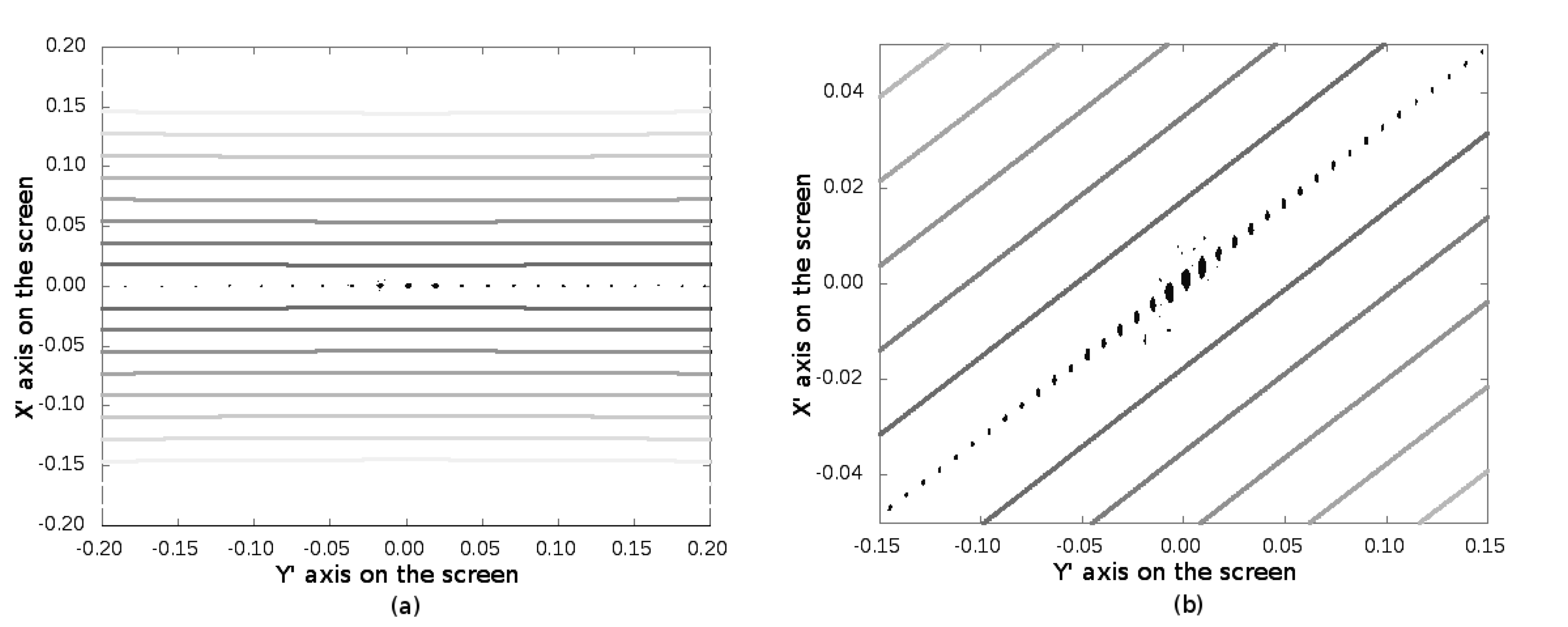}
\caption{Overlay of simulated and observed plots for variation of (a) only $\varphi$ and (b) only $\psi$. The values of the two angles are $-60.0^{\textup{o}}$ and $-18.2^{\textup{o}}$ respectively, in each case, the remaining two angles being zero.}
\label{phipsi}
\end{figure}
\vspace{-0.3cm}

\begin{figure}[h!]
\includegraphics[width=4.8in]{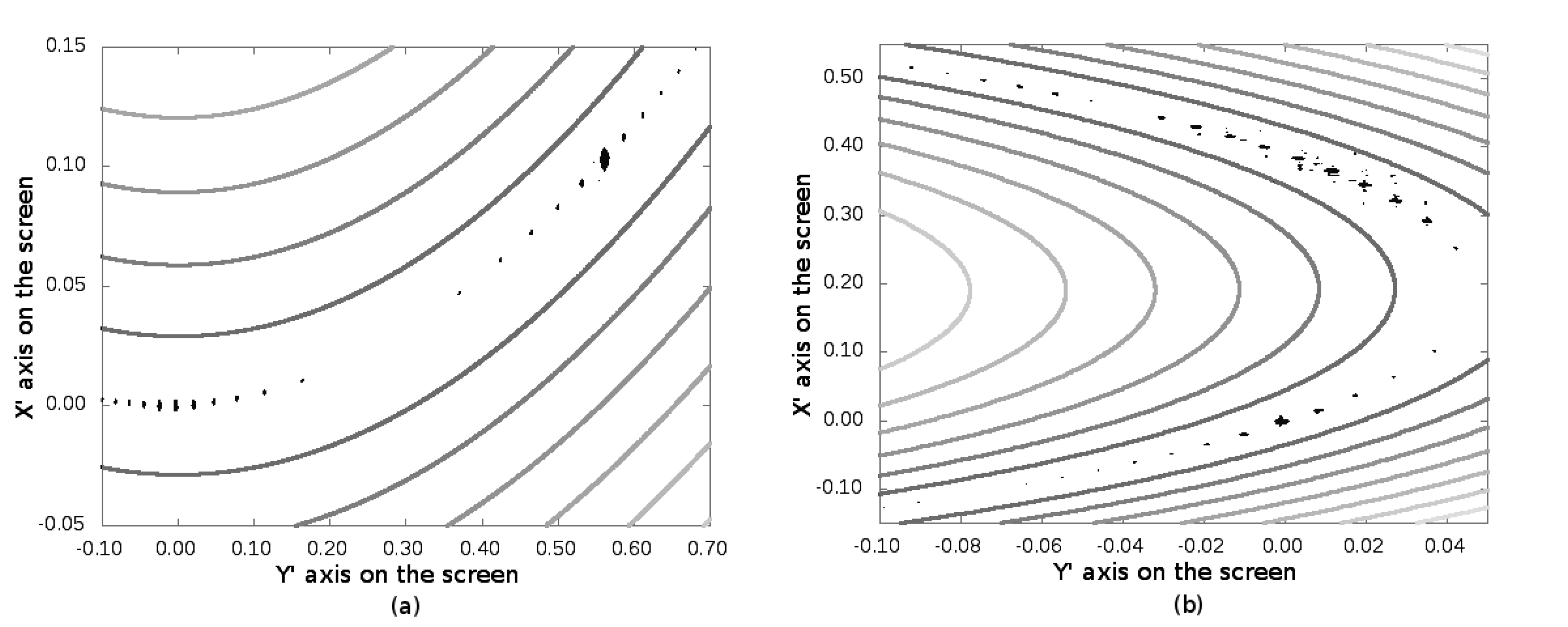}
\caption{Overlay of simulated and observed plots for variation of (a) $\theta + \varphi$ and (b) $\theta + \psi$. The values of the angles characterising these cases are respectively $\left( \theta=42.0^{\textup{o}},\varphi=72.6^{\textup{o}},\psi=0.0^{\textup{o}}\right)$ and $\left( \theta=82.4^{\textup{o}},\varphi=0.0^{\textup{o}},\psi=-14.8^{\textup{o}}\right)$.}
\label{thetaphithetapsi}
\end{figure}
\vspace{-0.3cm}

\begin{figure}[h!]
\includegraphics[width=4.8in]{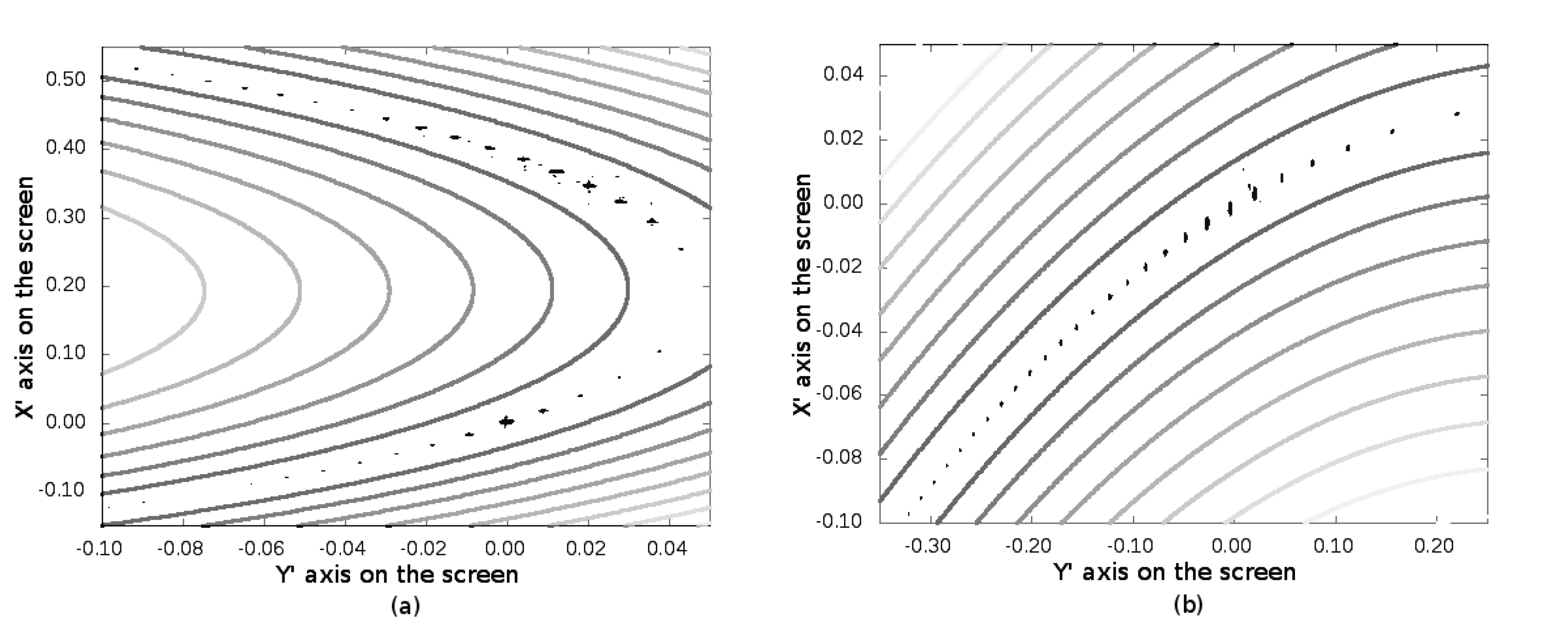}
\caption{Overlay of simulated and observed plots for variation of (a) $\varphi + \theta$ and (b) $\varphi + \psi$. The values of the angles characterising these cases are respectively $\left( \theta=82.4^{\textup{o}},\varphi=14.8^{\textup{o}},\psi=0.0^{\textup{o}}\right)$ and $\left( \theta=0.0^{\textup{o}},\varphi=74.8^{\textup{o}},\psi=-37.0^{\textup{o}}\right)$.}
\label{phithetaphipsi}
\end{figure}
\vspace{-0.3cm}

\newpage

\begin{figure}[h!]
\includegraphics[width=4.8in]{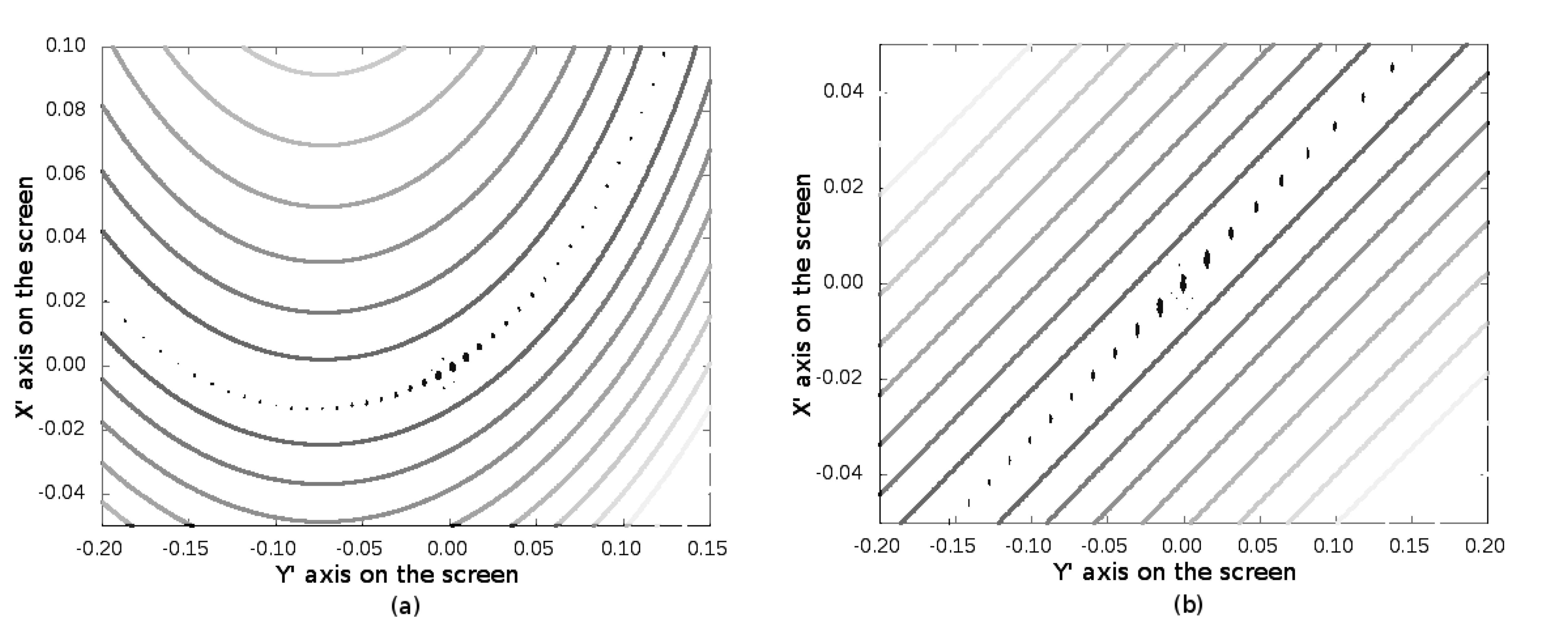}
\caption{Overlay of simulated and observed plots for variation of (a) $\psi + \theta$ and (b) $\psi + \varphi$. The values of the angles characterising these cases are respectively $\left( \theta=80.1^{\textup{o}},\varphi=0.0^{\textup{o}},\psi=-18.2^{\textup{o}}\right)$ and $\left( \theta=0.0^{\textup{o}},\varphi=70.0^{\textup{o}},\psi=-18.0^{\textup{o}}\right)$.}
\label{psithetapsiphi}
\end{figure}
\,
\twocolumngrid

\subsection{Variation of $\theta + \varphi$}
\vspace{-0.4cm}
\begin{subequations}
\setlength{\jot}{0.1cm}
\begin{align}
&Q = Q \left[ l\cos\left(\theta\right)-w\sin\left(\varphi\right)\sin\left(\theta\right), w\cos\left(\varphi\right), \right. \nonumber \\
&\left. \qquad l\sin\left(\theta\right)+w\sin\left(\varphi\right)\cos\left(\theta\right) \right], \label{tvq} \\[0.15cm]
&\beta_{l} = kL \left\lbrace \sin\left(\theta\right)\left[1-\cos\left(\alpha_{z}\right)\right] - \cos\left(\theta\right)\cos\left(\alpha_{x}\right) \right\rbrace, \label{tvl} \\[0.15cm]
&\beta_{w} = kW \left\lbrace \sin\left(\varphi\right)\left( \cos\left(\theta\right)\left[1-\cos\left(\alpha_{z}\right)\right]+\sin\left(\theta\right)\cos\left(\alpha_{x}\right) \right) \right. \nonumber \\
&\left. \qquad - \cos\left(\varphi\right)\cos\left(\alpha_{y}\right) \right\rbrace. \label{tvw}
\end{align}
\end{subequations}
The equation of locus is same as that for the variation of only $\theta$, Eq.~\eqref{te}.
\vspace{-0.3cm}

\subsection{Variation of $\theta + \psi$}
\begin{subequations}
\setlength{\jot}{0.1cm}
\begin{align}
&Q = Q \left[ l\cos\left(\psi\right)\cos\left(\theta\right)-w\sin\left(\psi\right)\cos\left(\theta\right), \right. \nonumber \\
&\left. \qquad l\sin\left(\psi\right)+w\cos\left(\psi\right), \right. \nonumber \\
&\left. \qquad l\cos\left(\psi\right)\sin\left(\theta\right)-w\sin\left(\psi\right)\sin\left(\theta\right) \right], \label{tsq} \\[0.15cm]
&\beta_{l} = kL \left\lbrace \cos\left(\psi\right)\left( \sin\left(\theta\right)\left[1-\cos\left(\alpha_{z}\right)\right] -\cos\left(\theta\right)\cos\left(\alpha_{x}\right) \right) \right. \nonumber \\
&\left. \qquad - \sin\left(\psi\right)\cos\left(\alpha_{y}\right) \right\rbrace, \label{tsl} \\[0.15cm]
&\beta_{w} = kW \left\lbrace -\sin\left(\psi\right)\left( \sin\left(\theta\right)\left[1-\cos\left(\alpha_{z}\right)\right]-\cos\left(\theta\right)\cos\left(\alpha_{x}\right) \right) \right. \nonumber \\
&\left. \qquad - \cos\left(\psi\right)\cos\left(\alpha_{y}\right) \right\rbrace. \label{tsw}
\end{align}
\end{subequations}
The equation of locus is:
\begin{align}
& \left\lbrace \left[X^{\prime}-L\cos\left(\psi\right)\cos\left(\theta\right)\right]^{2} + \left[Y^{\prime}-L\sin\left(\theta\right)\right]^{2}\right. \nonumber \\
&\left. \quad + \left[D^{\prime}-L\cos\left(\psi\right)\sin\left(\theta\right)\right]^{2} \right\rbrace^{1/2} + 2L\cos\left(\psi\right)\sin\left(\theta\right) \nonumber \\
&= \left\lbrace \left[X^{\prime}+L\cos\left(\psi\right)\cos\left(\theta\right)\right]^{2} + \left[Y^{\prime}+L\sin\left(\theta\right)\right]^{2}\right. \nonumber \\
&\left. \quad + \left[D^{\prime}+L\cos\left(\psi\right)\sin\left(\theta\right)\right]^{2} \right\rbrace^{1/2}. \label{tse}
\end{align}
\vspace{-0.8cm}

\subsection{Variation of $\varphi + \theta$}
\begin{subequations}
\setlength{\jot}{0.1cm}
\begin{align}
&Q = Q \left[ l\cos\left(\theta\right), -l\sin\left(\theta\right)\sin\left(\varphi\right)+w\cos\left(\varphi\right),\right. \nonumber \\
&\left. \qquad l\sin\left(\theta\right)\cos\left(\varphi\right)+w\sin\left(\varphi\right) \right], \label{vtq} \\[0.15cm]
&\beta_{l} = kL \left\lbrace \sin\left(\theta\right)\left( \cos\left(\varphi\right)\left[1-\cos\left(\alpha_{z}\right)\right]+\sin\left(\varphi\right)\cos\left(\alpha_{y}\right) \right) \right. \nonumber \\
&\left. \qquad -\cos\left(\theta\right)\cos\left(\alpha_{x}\right) \right\rbrace, \label{vtl} \\[0.15cm]
&\beta_{w} = kW \left\lbrace \sin\left(\varphi\right)\left[1-\cos\left(\alpha_{z}\right)\right] - \cos\left(\varphi\right)\cos\left(\alpha_{y}\right) \right\rbrace. \label{vtw}
\end{align}
\end{subequations}
The equation of locus is:
\begin{align}
& \left\lbrace \left[X^{\prime}-L\cos\left(\theta\right)\right]^{2} + \left[Y^{\prime}-L\sin\left(\theta\right)\sin\left(\varphi\right)\right]^{2} \right. \nonumber \\
&\left. \quad + \left[D^{\prime}-L\sin\left(\theta\right)\cos\left(\varphi\right)\right]^{2} \right\rbrace^{1/2} + 2L\sin\left(\theta\right)\cos\left(\varphi\right) \nonumber \\
&= \left\lbrace \left[X^{\prime}+L\cos\left(\theta\right)\right]^{2} + \left[Y^{\prime}+L\sin\left(\theta\right)\sin\left(\varphi\right)\right]^{2} \right. \nonumber \\
&\left. \quad + \left[D^{\prime}+L\sin\left(\theta\right)\cos\left(\varphi\right)\right]^{2} \right\rbrace^{1/2}. \label{vte}
\end{align}

\subsection{Variation of $\varphi + \psi$}
\begin{subequations}
\setlength{\jot}{0.1cm}
\begin{align}
&Q = Q \left[l\cos\left(\psi\right)-w\sin\left(\psi\right), \right. \nonumber \\
&\left. \qquad l\sin\left(\psi\right)\cos\left(\varphi\right)+w\cos\left(\psi\right)\cos\left(\varphi\right), \right. \nonumber \\
&\left. \qquad l\sin\left(\psi\right)\sin\left(\varphi\right)+w\cos\left(\psi\right)\sin\left(\varphi\right) \right], \label{vsq} \\[0.15cm]
&\beta_{l} = kL \left\lbrace \sin\left(\psi\right)\left( \sin\left(\varphi\right)\left[1-\cos\left(\alpha_{z}\right)\right]-\cos\left(\varphi\right)\cos\left(\alpha_{y}\right) \right) \right. \nonumber \\
&\left. \qquad - \cos\left(\psi\right)\cos\left(\alpha_{x}\right) \right\rbrace, \label{vsl} \\[0.15cm]
&\beta_{w} = kW \left\lbrace \cos\left(\psi\right)\left( \sin\left(\varphi\right)\left[1-\cos\left(\alpha_{z}\right)\right]-\cos\left(\varphi\right)\cos\left(\alpha_{y}\right) \right) \right. \nonumber \\
&\left. \qquad + \sin\left(\psi\right)\cos\left(\alpha_{x}\right) \right\rbrace . \label{vsw}
\end{align}
\end{subequations}
The equation of locus is:
\begin{align}
& \left\lbrace \left[X^{\prime}-L\cos\left(\psi\right)\right]^{2} + \left[Y^{\prime}-L\sin\left(\psi\right)\cos\left(\varphi\right)\right]^{2} \right. \nonumber \\
&\left. \quad + \left[D^{\prime}-L\sin\left(\psi\right)\sin\left(\varphi\right)\right]^{2} \right\rbrace^{1/2} + 2 L \sin \left( \psi \right)\sin \left( \varphi \right) \nonumber \\
&= \left\lbrace \left[X^{\prime}+L\cos\left(\psi\right)\right]^{2} + \left[Y^{\prime}+L\sin\left(\psi\right)\cos\left(\varphi\right)\right]^{2} \right. \nonumber \\
&\left. \quad + \left[D^{\prime}+L\sin\left(\psi\right)\sin\left(\varphi\right)\right]^{2} \right\rbrace^{1/2}. \label{vse}
\end{align}

\subsection{Variation of $\psi + \theta$}
\begin{subequations}
\setlength{\jot}{0.1cm}
\begin{align}
&Q = Q \left[l\cos\left(\theta\right)\cos\left(\psi\right)-w\sin\left(\psi\right),\right. \nonumber \\
&\left. \qquad l\cos\left(\theta\right)\sin\left(\psi\right)+w\cos\left(\psi\right),l\sin\left(\theta\right) \right], \label{stq} \\[0.15cm]
&\beta_{l} = kL \left\lbrace \sin\left(\theta\right)\left[1-\cos\left(\alpha_{z}\right)\right] \right. \nonumber \\
&\left. \qquad -\cos\left(\theta\right)\left(\cos\left(\psi\right)\cos\left(\alpha_{x}\right)+ \sin\left(\psi\right)\cos\left(\alpha_{y}\right)\right) \right\rbrace, \label{stl} \\[0.15cm]
&\beta_{w} = kW \left\lbrace \sin\left(\psi\right)\cos\left(\alpha_{x}\right)- \cos\left(\psi\right)\cos\left(\alpha_{y}\right) \right\rbrace. \label{stw}
\end{align}
\end{subequations}
The equation of locus is:
\begin{align}
& \left\lbrace \left[X^{\prime}-L\cos\left(\theta\right)\cos\left(\psi\right)\right]^{2} + \left[Y^{\prime}-L\cos\left(\theta\right)\sin\left(\psi\right)\right]^{2} \right. \nonumber \\
&\left. \quad + \left[D^{\prime}-L\sin\left(\theta\right)\right]^{2} \right\rbrace^{1/2} + 2L\sin\left(\theta\right) \nonumber \\
&= \left\lbrace \left[X^{\prime}+L\cos\left(\theta\right)\cos\left(\psi\right)\right]^{2} + \left[Y^{\prime}+L\cos\left(\theta\right)\sin\left(\psi\right)\right]^{2} \right. \nonumber \\
&\left. \quad + \left[D^{\prime}+L\sin\left(\theta\right)\right]^{2} \right\rbrace^{1/2}. \label{ste}
\end{align}
\vspace{-1.0cm}

\subsection{Variation of $\psi + \varphi$}
\label{results-J}
\vspace{-0.5cm}
\begin{subequations}
\setlength{\jot}{0.1cm}
\begin{align}
&Q = Q \left[l\cos\left(\psi\right)-w\cos\left(\varphi\right)\sin\left(\psi\right),\right. \nonumber \\
&\left. \qquad l\sin\left(\psi\right)+w\cos\left(\varphi\right)\cos\left(\psi\right),w\sin\left(\varphi\right) \right], \label{svq} \\[0.15cm]
&\beta_{l} = kL \left\lbrace \cos\left(\psi\right)\cos\left(\alpha_{x}\right) + \sin\left(\psi\right)\cos\left(\alpha_{y}\right)\right\rbrace, \label{svl} \\[0.15cm]
&\beta_{w} = kW \left\lbrace \sin\left(\varphi\right)\left[1-\cos\left(\alpha_{z}\right)\right] \right. \nonumber \\
&\left. \qquad -\cos\left(\varphi\right)\left(\sin\left(\psi\right)\cos\left(\alpha_{x}\right)- \cos\left(\psi\right)\cos\left(\alpha_{y}\right)\right) \right\rbrace. \label{svw}
\end{align}
\end{subequations}
The equation of locus is same as that for the variation of only $\psi$, Eq.~\eqref{se}.
\vspace{-0.3cm}

\section{Analysis and Discussion} \label{discussion}
\vspace{-0.3cm}
Figures \ref{nitheta}--\ref{psithetapsiphi} illustrate the agreement between theory and experiment for all ten cases. Even though still photographs have been presented as observations, we have also qualitatively studied the locus with change in angle using numerical simulations.\cite{completederivations} Intensity plots are generated over a range of angles for each case and stitched together to form movies\cite{youtube} using the software application \textsc{mencoder}.\cite{mencoder} In this section, we analyse some peculiar features of the locus for each case, simplifying the equations wherever possible. In cases where simplification is not straightforward, we take a heuristic approach to understand the locus.

For the case of normal incidence $(NI)$, the equation of locus [Eq.~\eqref{nie}] can be simplified to  $X^{\prime}=0$, which is the $Y^\prime$ axis on the screen. For a slit normal to the incident light we also expect the locus of points equidistant from the ends to be a horizontal line. This is due to the symmetric placement of the points $A$ and $B$ about the origin $O$ in a plane parallel to that of the screen.

For variation of only $\theta$, the equation of locus [Eq.~\eqref{te}] can be simplified to that of a conic section, a second degree curve.\cite{phadke} The type of conic section is determined by the value of~$\theta$.

The equation of locus for variation of only $\varphi$ is the same as that for normal incidence [Eq.~\eqref{nie}], because changing $\varphi$ does not alter the effective length of the slit. However, the decreasing effective width of the slit with angle $\varphi$ results in a larger separation between adjacent diffraction maxima [see Fig.~\hyperref[phipsi]{\ref{phipsi}(a)}] as compared to the case of normal incidence [Fig.~\hyperref[nitheta]{\ref{nitheta}(a)}]. Also, one can count nine maxima in the $-Y^\prime$ direction, as opposed to eleven in the $+Y^\prime$ direction, because of the asymmetry of the situation.

For variation of only $\psi$, as the effective length of the slit is the same as that for normal incidence, we expect the locus to be a line with a different slope, owing to the changes in the coordinates of $A$ and~$B$. After simplifying the equation of locus [Eq.~\eqref{se}], we obtain $X^{\prime}=-\tan\left(\psi\right)Y^{\prime}$ which is the equation of a line passing through the origin, with a negative slope of $-\tan\left(\psi\right)$. In our case $\psi$ is negative and the slope of the observed locus [Fig.~\hyperref[phipsi]{\ref{phipsi}(b)}] is accordingly positive.

For variation of $\theta + \varphi$, fixing $\theta$ and varying $\varphi$ changes neither the expressions for $A$ and $B$ nor the effective length of the slit. Hence the equation of locus is exactly the same as that for varying only~$\theta$. Increasing the tilt by angle $\varphi$ causes the reflected diffraction maxima to move along the locus and come closer to the transmitted maxima.\cite{youtube} For the given values of $\theta$ and $\varphi$ one can locate the reflected 0th-order maximum (RZOM) in Fig.~\hyperref[thetaphithetapsi]{\ref{thetaphithetapsi}(a)} at approximately $\left(0.55, 0.1\right)$, which would slowly approach the transmitted 0th-order maximum (TZOM) at $\left(0, 0\right)$ as $\varphi \rightarrow 90^{\textup{o}}$.

Fixing angle $\theta$ and varying the angle $\psi$ for the variation of $\theta + \psi$ should not change the $Y^{\prime}$ coordinate of the RZOM as the plane of the grating remains unchanged. As expected, the observed $Y^\prime$ coordinates of the TZOM and RZOM in Fig.~\hyperref[thetaphithetapsi]{\ref{thetaphithetapsi}(b)} lie within 1~cm of each other.

Observe that the plots for variation of $\theta + \psi$ [Fig.~\hyperref[thetaphithetapsi]{\ref{thetaphithetapsi}(b)}] and $\varphi + \theta$ [Fig.~\hyperref[phithetaphipsi]{\ref{phithetaphipsi}(a)}] are apparently similar, even though the equations of the locus for the two cases are different. This is because for the large angle of $\theta = 82.4^{\textup{o}}$, the same value of the other two angles orients the grating in a similar manner. One could think of the limiting case of angle $\theta = 90^{\textup{o}}$ and the angles $\psi$ and $\varphi$ to have the same value, for which the final orientation of the grating is exactly the same and the diffraction patterns are identical.

The two cases involving the variation of $\varphi + \psi$ and $\psi + \theta$ are related to each other in a manner similar to the earlier two. If $\psi = 90^{\textup{o}}$ and the angles $\theta$ and $\varphi$ have the same value, the diffraction patterns are identical. The diffraction pattern for variation of $\psi + \theta$ may be visualized as being that of angle $\theta$, which has been rotated by an angle $\psi$ [Fig.~\hyperref[psithetapsiphi]{\ref{psithetapsiphi}(a)}].

Because angle $\varphi$ changes neither the expressions for $A$ and $B$ nor the effective length of the slit, the equation of locus for variation of $\psi + \varphi$ is same as that for angle $\psi$ [Eq.~\eqref{se}]. The diffraction pattern [Fig.~\hyperref[psithetapsiphi]{\ref{psithetapsiphi}(b)}] can be visualized as being that of angle $\varphi$, which has been rotated by an angle~$\psi$.

\vspace{-0.3cm}

\section{Conclusion} \label{conclusion}
\vspace{-0.2cm}
Using the Fresnel-Kirchhoff formula, we have seen that the intensity distribution for all the ten cases has the same form [Eq.~\eqref{grating}]. We have argued that the locus of diffraction maxima is suitable for quantitative observations across all ten cases. In this regard, it is important to note that the coordinates of the points $A$ and $B$ decide the equation of locus. In the cases $NI, \varphi, \psi$, and $\psi + \varphi$ where $A$ and $B$ lie in the $X$-$Y$ plane, the equation of locus can be simplified to that of a line (a first degree curve). Tilting the slit into an orientation other than these leads to an initial path difference between points $A$ and~$B$. This prevents the simplification of the equation of locus to that of a line. For these cases, the equation of locus is a second degree curve, which gives the diffraction pattern its curved shape.

We suggest that these novel cases of diffraction be presented as interesting educational detours from the case of normal incidence, both in the undergraduate classroom and in the laboratory. The relevant theory is tractable through use of the Fraunhofer diffraction limit and yet has something novel to offer. We have noted that identifying the coordinates of $Q$ itself is a test of one's visualization abilities. In an undergraduate laboratory, some of these cases may be presented as experimental activities which will help the students to develop conceptual understanding and foster their abilities and skills required for visualization, simulation, and image processing. Students can try these experiments with single or multiple slits using photosensitive screens to obtain information about the coordinates of maxima and compare their observations with simulations.\cite{completederivations}

\vspace{-0.3cm}
\begin{acknowledgments}
\vspace{-0.2cm}
This undergraduate student project was carried out under the National Initiative on Undergraduate Science (NIUS) Programme of HBCSE-TIFR, Mumbai, India. The authors are grateful to the authorities at HBCSE-TIFR, Mumbai and UM-DAE CEBS, Mumbai for supporting our project. They are thankful to the staff at NIUS Optics Laboratory for their help during the study.
\end{acknowledgments}


\begin{thebibliography}{5}
\bibitem{rect}In the literature, a distinction is made between a single slit (of indefinite length, but definite width $W$) and a rectangular aperture (of definite length $L$ and width $W$). Even though we study the rectangular aperture, we refer to it as the slit. Here, we consider a slit as a rectangular aperture with $L \gg W$.
\bibitem{arrangement}This arrangement of rings, which is symmetric about the middle $\theta$~ring, was arrived at by a set of repeated trials. However, determining the minimum number of inter-connected rings that can allow for all ten orientations can be an interesting group-theoretic problem.
\end{thebibliography}
\end{document}


\section*{Supplement for \emph{Novel cases of diffraction of light from a grating: Theory and experiment}}
\vspace{-0.3cm}

This supplement consists of two sections. The first section provides a sample derivation of the diffraction intensity on the screen. Complete derivations of the diffraction intensity for all 10 cases of grating orientation, are provided in the Supplemental Material at \url{https://dx.doi.org/10.1119/1.4737854} as well as on our website at \url{https://sites.google.com/site/noveldiffractionnrjasrbk/derivations}.

The second section describes in detail how the experimental results were checked against theoretical plots for the locus of the diffraction maxima by using image processing techniques. This is also available in the Supplemental Material as well as on our website at \url{https://sites.google.com/site/noveldiffractionnrjasrbk/processing}.

\vspace{-0.6cm}

\section*{A. Derivation for variation of $\varphi+\psi$}
\vspace{-0.3cm}

As a sample derivation, in this section we look at the case of $\varphi + \psi$, corresponding to the results described in Sec. V H of the paper.

\begin{figure}[h!]
\includegraphics[width=6.1in]{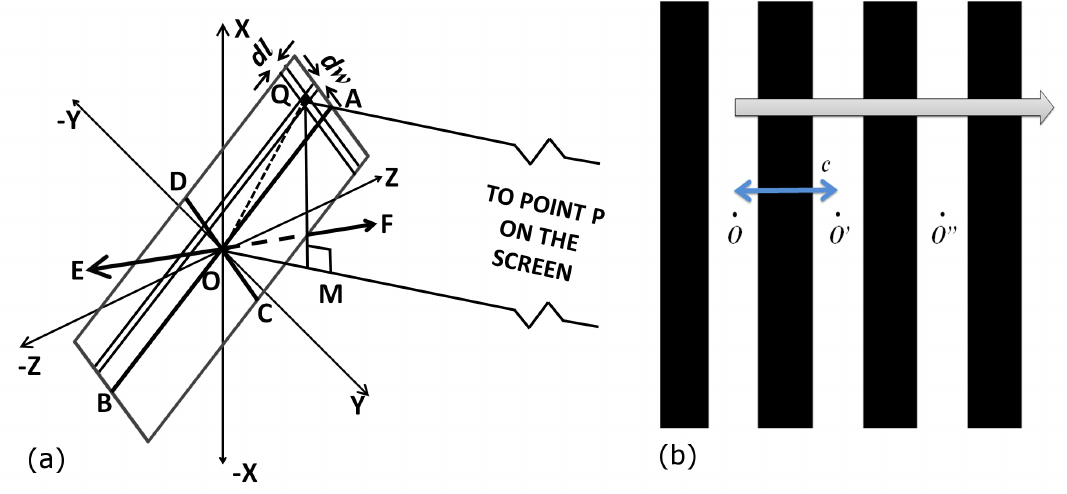}
\caption{(a) Schematic of the geometry used in the integral of the Fresnel-Kirchhoff formula. Depending on orientation, the co-ordinates of $Q$ will be functions of the two angles involved. (b) Schematic of the grating showing the origin $O$, and the centres of the adjacent slits $O^{\prime}$, $O^{\prime\prime}$ and so on.}
\label{derivation}
\end{figure}
\vspace{-0.3cm}

Let us first obtain the intensity distribution due to a single slit. Consider a point Q on the surface of the slit with co-ordinates $(Q_{x}\,, Q_{y}\,, Q_{z}\,)$, as illustrated in Fig. 1 (a). In terms of the angles $\varphi$ and $\psi$ 
\begin{equation}
Q= Q \left[lcos\left(\psi\right)-wsin\left(\psi\right), lsin\left(\psi\right)cos\left(\varphi\right)+wcos\left(\psi\right)cos\left(\varphi\right), lsin\left(\psi\right)sin\left(\varphi\right)+wcos\left(\psi\right)sin\left(\varphi\right)\right].
\end{equation}
With the origin $O=O\left[ 0, 0, 0 \right]$ and unit vetors $\hat{i}, \hat{j}$ and $\hat{k}$ along the X, Y and Z axis respectively,
\begin{equation}
\mathbf{OQ} = Q_{x}\,\hat{i}+Q_{y}\,\hat{j}+Q_{z}\,\hat{k}.
\end{equation}
For a point $P=P\left[X^{\prime},Y^{\prime},D^{\prime}\right]$ on the screen,
\begin{equation}
\mathbf{OP}=R\,cos\left(\alpha_{x}\right)\hat{i} + R\,cos\left(\alpha_{y}\right)\hat{j} + R\,cos\left(\alpha_{z}\right)\hat{k},
\end{equation}
where
\begin{align}
&R=\sqrt{X^{\prime2}+Y^{\prime2}+D^{\prime2}}, \nonumber \\
&cos\left(\alpha_{x}\right)=\frac{X^{\prime}}{R}, \nonumber \\
&cos\left(\alpha_{y}\right)=\frac{Y^{\prime}}{R}, \nonumber \\
&cos\left(\alpha_{z}\right)=\frac{D^{\prime}}{R}. \nonumber
\end{align}
$OM$ is given by the norm of the projection of $\mathbf{OQ}$ along $\mathbf{OP}$
\begin{align}
OM&= Q_{x}\,cos\left(\alpha_{x}\right)+Q_{y}\,cos\left(\alpha_{y}\right)+Q_{z}\,cos\left(\alpha_{z}\right), \nonumber \\
&=\left[lcos\left(\psi\right)-wsin\left(\psi\right)\right]cos\left(\alpha_{x}\right) \nonumber\\
&+\left[lsin\left(\psi\right)cos\left(\varphi\right)+wcos\left(\psi\right)cos\left(\varphi\right)\right]
cos\left(\alpha_{y}\right)\nonumber\\
&+\left[lsin\left(\psi\right)sin\left(\varphi\right)+wcos\left(\psi\right)sin\left(\varphi\right)\right]
cos\left(\alpha_{z}\right).
\end{align}
The total path difference is
\begin{align}
QP-OP&=Q_{z}-OM, \nonumber \\
&=lsin\left(\psi\right)sin\left(\varphi\right)+wcos\left(\psi\right)sin\left(\varphi\right)- \left[lcos\left(\psi\right)-wsin\left(\psi\right)\right]cos\left(\alpha_{x}\right) \nonumber\\
&-\left[lsin\left(\psi\right)cos\left(\varphi\right)+wcos\left(\psi\right)cos\left(\varphi\right)\right]
cos\left(\alpha_{y}\right) \nonumber\\
&-\left[lsin\left(\psi\right)sin\left(\varphi\right)+wcos\left(\psi\right)sin\left(\varphi\right)\right]
cos\left(\alpha_{z}\right),
\end{align}
which can be written as (after grouping the terms in $l$ and $w$ together)
\begin{align}
Q_{z}-OM&= l \left\lbrace sin\left(\psi\right)\left( sin\left(\varphi\right)\left[1-cos\left(\alpha_{z}\right)\right]- cos\left(\varphi\right)cos\left(\alpha_{y}\right) \right) - cos\left(\psi\right)cos\left(\alpha_{x}\right) \right\rbrace \nonumber\\
&+w \left\lbrace cos\left(\psi\right)\left( sin\left(\varphi\right)\left[1-cos\left(\alpha_{z}\right)\right]- cos\left(\varphi\right)cos\left(\alpha_{y}\right) \right) + sin\left(\psi\right)cos\left(\alpha_{x}\right) \right\rbrace.
\end{align}
The total optical disturbance $(U)$ travelling along the direction of point P, as given by the Fresnel-Kirchhoff formula is
\begin{equation}
U =H \!\int_{area} \!\!\! dw \, dl \,exp \left[ \iota k \left\lbrace Q_{z}-OM \right\rbrace \right], \label{amp}
\end{equation}
where $k$ is the wavenumber and $H$ is a constant. Therefore,
\begin{align}
U &=H \!\int_{-L}^{L} \!\!\! dl \,exp \left[ \iota kl \left\lbrace sin\left(\psi\right)\left( sin\left(\varphi\right)\left[1-cos\left(\alpha_{z}\right)\right]- cos\left(\varphi\right)cos\left(\alpha_{y}\right) \right) - cos\left(\psi\right)cos\left(\alpha_{x}\right)\right\rbrace \right] \nonumber \\
& \times \int_{-W}^{W} \!\!\! dw \,exp \left[ \iota kw \left\lbrace cos\left(\psi\right)\left( sin\left(\varphi\right)\left[1-cos\left(\alpha_{z}\right)\right]- cos\left(\varphi\right)cos\left(\alpha_{y}\right) \right) + sin\left(\psi\right)cos\left(\alpha_{x}\right)  \right\rbrace \right].
\label{fkf}
\end{align}After integrating and rearranging, we obtain,
\begin{equation}
U = H^{\prime} \left( \frac{sin \left( \beta_{l} \right)}{\beta_{l}} \right) \left( \frac{sin \left( \beta_{w} \right)}{\beta_{w}} \right) \label{rect},
\end{equation}
where,
\begin{subequations}
\begin{align}
H^{\prime}&= H\left(2L\right)\left(2W\right) \label{hprime}, \\
\beta_{l} &= kL\left\lbrace sin\left(\psi\right)\left( sin\left(\varphi\right)\left[1-cos\left(\alpha_{z}\right)\right]- cos\left(\varphi\right)cos\left(\alpha_{y}\right) \right) - cos\left(\psi\right)cos\left(\alpha_{x}\right)\right\rbrace \label{betal}, \\
\beta_{w} &= kW\left\lbrace cos\left(\psi\right)\left( sin\left(\varphi\right)\left[1-cos\left(\alpha_{z}\right)\right]- cos\left(\varphi\right)cos\left(\alpha_{y}\right) \right) + sin\left(\psi\right)cos\left(\alpha_{x}\right) \right\rbrace \label{betaw}.
\end{align}
\end{subequations}
Now let us consider the case of a grating with grating period $c$, of which $N$ slits out of the total have been illuminated. The amplitude distribution due to a grating $\left(U_{G}\right)$ is given by the addition of these $N$ number of amplitudes. Consider the centre $O^{\prime}$ of the adjacent slit with co-ordinates $\left(c_{x}\,,c_{y}\,,c_{z}\,\right)$ as illustrated in Fig. 1 (b). In terms of the angles $\phi$ and $\psi$
\begin{align}
O^{\prime}&=O^{\prime}\left[-c\, sin\left(\psi\right), c\,cos\left(\psi\right)cos\left(\varphi\right), c\, cos\left(\psi\right)sin\left(\varphi\right)\right], \\
\mathbf{OO^{\prime}}&=c_{x}\,\hat{i} + c_{y}\,\hat{j} + c_{z}\,\hat{k}.
\end{align}
Similar to the definition of $OM$, we define $OM^{\prime}$ (not shown in the Fig. 1 (b)) as the norm of the projection of $\mathbf{OO^{\prime}}$ on $\mathbf{OP}$
\begin{align}
OM^{\prime}&= c_{x}\, cos\left(\alpha_{x}\right)+c_{y}\, cos\left(\alpha_{y}\right)+c_{z}\, cos\left(\alpha_{z}\right), \nonumber \\
&=-c\, sin\left(\psi\right)cos\left(\alpha_{x}\right) + c\,cos\left(\psi\right)cos\left(\varphi\right)cos\left(\alpha_{y}\right)+ 
c\, cos\left(\psi\right)sin\left(\varphi\right)cos\left(\alpha_{z}\right).
\end{align}
The path difference between adjacent slits is given by
\begin{equation}
c_{z}-OM^{\prime}= c\, cos\left(\psi\right)sin\left(\varphi\right)+                                                                              c\, sin\left(\psi\right)cos\left(\alpha_{x}\right) - c\,cos\left(\psi\right)cos\left(\varphi\right)cos\left(\alpha_{y}\right)- 
c\, cos\left(\psi\right)sin\left(\varphi\right)cos\left(\alpha_{z}\right),
\nonumber
\end{equation}
and the phase difference is
\begin{equation}
kc\left\lbrace cos\left(\psi\right)\left( sin\left(\varphi\right)\left[1-cos\left(\alpha_{z}\right)\right]- cos\left(\varphi\right)cos\left(\alpha_{y}\right) \right) + sin\left(\psi\right)cos\left(\alpha_{x}\right) \right\rbrace = \beta_{c}. \label{betac}
\end{equation}
As we sum over successive slits located at $O^{\prime}$, $O^{\prime\prime}$ and so on, the net amplitude $\left(U_{G}\right)$ due to a grating is given in terms of $U$ as 
\begin{align}
U_{G}&=U + Ue^{\iota\beta_{c}}+ Ue^{\iota2\beta_{c}}+ \dots + Ue^{\iota N \beta_{c}}, \nonumber \\
&=U\left(1 + e^{\iota\beta_{c}}+ e^{\iota2\beta_{c}}+ \dots + e^{\iota N \beta_{c}} \right), \nonumber \\
\implies U_{G}&=U\left(\frac{1- e^{\iota N \beta_{c}}}{1- e^{\iota\beta_{c}}}\right).
\end{align}
The intensity distribution $\left(I\right)$ is given by multiplying $U_{G}$ by its complex conjugate
\begin{align}
I&=U_{G}^{*}U_{G}, \nonumber \\
&=U^{2}\left(\frac{1- e^{-\iota N \beta_{c}}}{1- e^{-\iota\beta_{c}}}\right) \left(\frac{1- e^{\iota N \beta_{c}}}{1- e^{\iota\beta_{c}}}\right), \nonumber \\
&=U^{2}\left(\frac{sin\left(\frac{N\beta_{c}}{2}\right)}{sin\left(\frac{\beta_{c}}{2}\right)}\right)^{2}, \nonumber \\
\implies I &= \left( H^{\prime} \right)^{2} \left( \frac{sin \left( \beta_{l} \right)}{\beta_{l}} \right)^{2} \left( \frac{sin \left( \beta_{w} \right)}{\beta_{w}} \right)^{2} \left(\frac{sin\left(\frac{N\beta_{c}}{2}\right)}{sin\left(\frac{\beta_{c}}{2}\right)}\right)^{2}. \label{grating}
\end{align}
as required, with the expressions for $H^{\prime}$, $\beta_{l}$, $\beta_{w}$ and $\beta_{c}$ given by Eqs. \ref{hprime}, \ref{betal}, \ref{betaw} and \ref{betac} respectively.

\section*{B. Details of Image Processing}
\begin{enumerate}
\item Here, we will go through the steps of image processing and analysis done for the sample case of $\varphi + \psi$. First, capture the diffraction pattern that you want to analyse. The captured image for the sample case is shown in Fig. \ref{raw}
\begin{figure}[h!]
\centering
\includegraphics[width=4.0 in]{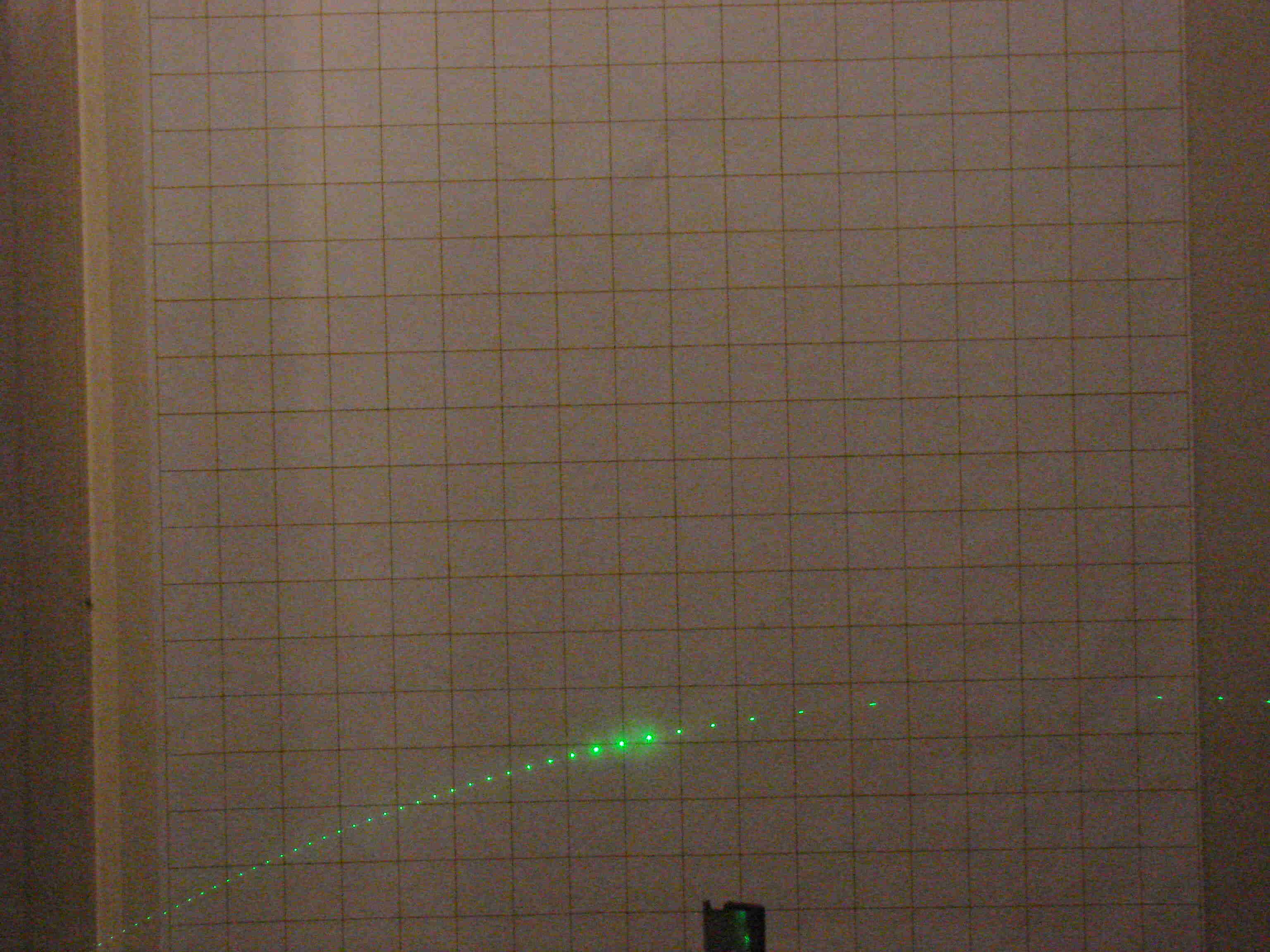}
\caption{Image of the screen captured during the experiment, for the case of $\varphi + \psi$.}
\label{raw}
\end{figure}

\item Crop the above image for the part that you are interested in, shown in Fig. \ref{crop}
\begin{figure}[h!]
\centering
\includegraphics[width=4.0 in]{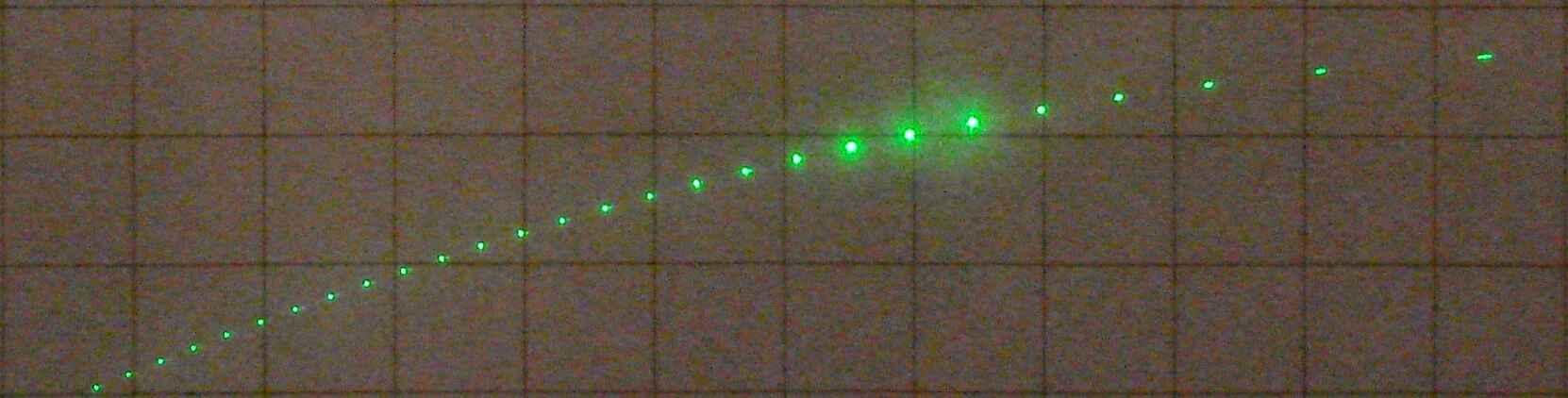}
\caption{The image in Fig.~\ref{raw} after being cropped.}
\label{crop}
\end{figure}

\item Separately, by running the program for locus in Octave3.2, generate the contour plot of the function $\left(AP+2Z\right)-BP$. The plot generated for this case is shown in Fig. \ref{locus}
\begin{figure}[h!]
\centering
\includegraphics[width=5.0 in]{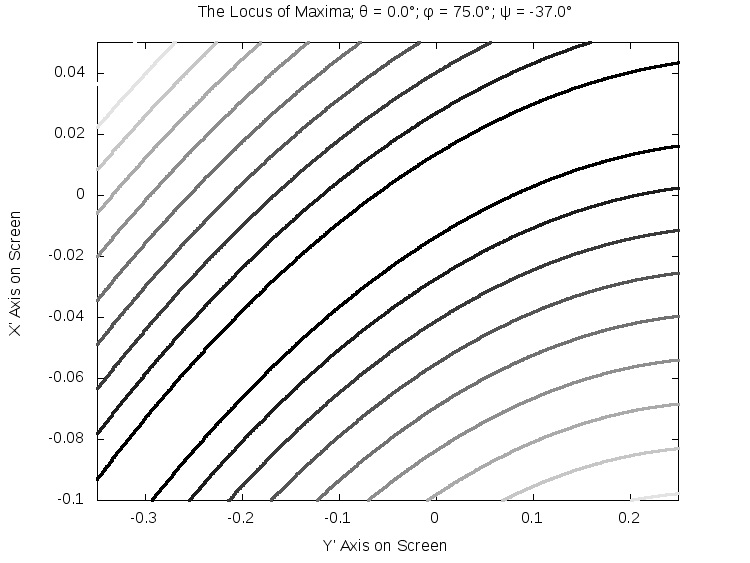}
\caption{The simulated contour plot of the function $\left(AP+2Z\right)-BP$ for the case $\varphi + \psi$.}
\label{locus}
\end{figure}

\item Now open GIMP and create a new blank background of the same dimensions as the simulated locus plot (Go to \textit{File$\rightarrow$New} or press \textit{Ctrl+N}). In this case, the simulated plot is $750$ pixels $\times 563$ pixels (see Fig. \ref{new}) in landscape mode because the image is such. You can see two other windows on the left and right. For the sake of future reference, we will call the one on the left as the toolbox window, and the one on the right as the layers window.
\begin{figure}[h!]
\centering
\includegraphics[width=7.1 in]{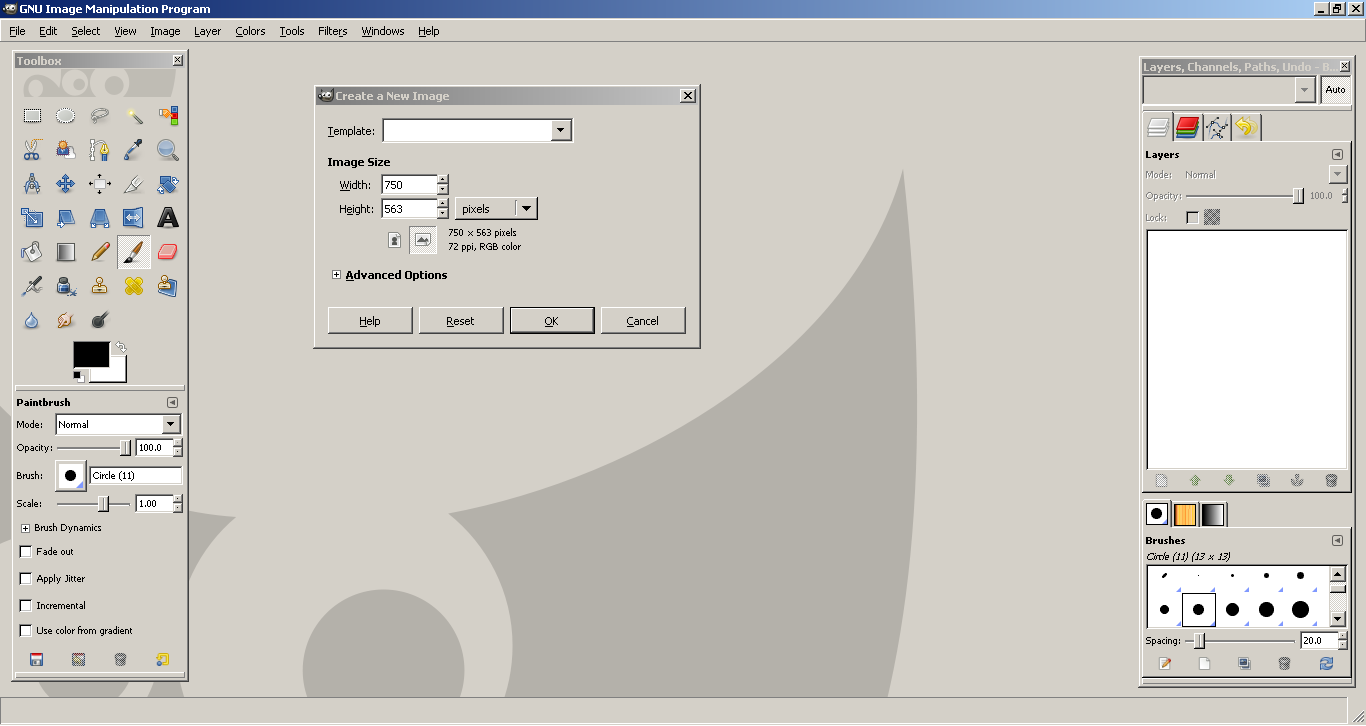}
\caption{A screenshot showing the making of a background of specific dimensions.}
\label{new}
\end{figure}

\item Next, open (Go to \textit{File$\rightarrow$Open} or press \textit{Ctrl+O}) the simulated plot, copy it (Go to \textit{Edit$\rightarrow$Copy} or press \textit{Ctrl+C}) and paste (Go to \textit{Edit$\rightarrow$paste} or press \textit{Ctrl+V}) it onto the background (see Fig. \ref{pastesimulation}). Notice (in Fig. \ref{pastesimulation}) that on layers window, the entry \textit{Floating Selection (Pasted Layer)} has been selected by a right click. Click on the new layer option. By doing this, we are treating this pasted layer as a new layer, which can be edited separately from the background - you will see why this is important in a while. You may name the pasted layer as `simulation'
\begin{figure}[h!]
\centering
\includegraphics[width=7.1 in]{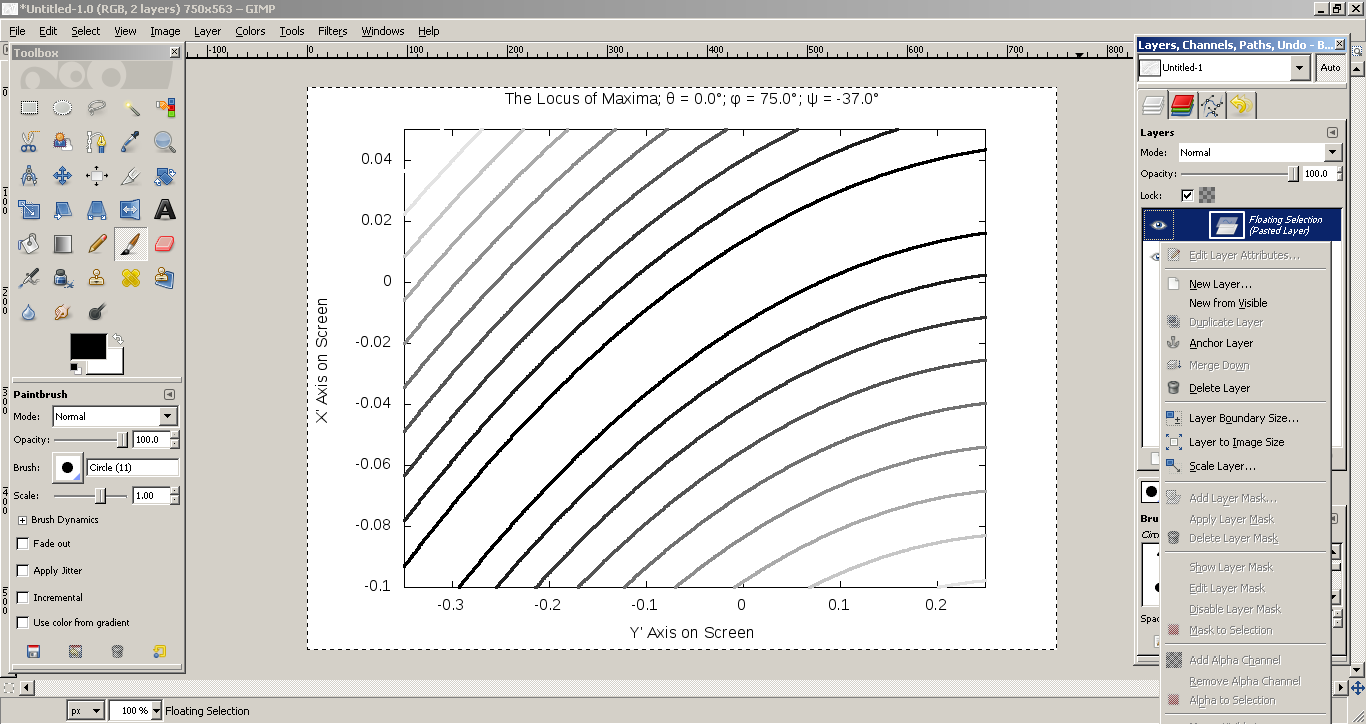}
\caption{A screenshot showing the contour plot as a new `simulation' layer.}
\label{pastesimulation}
\end{figure}

\item Do the same as above for the cropped image. Name this pasted layer as `experiment'. Now if you look at the Layers window (see Fig. \ref{layerwindow}), you should have a background, a layer named simulation and a layer named experiment. Note that you can change the opacity of each layer separately from here. Try it out! You will need it in the next step. \label{step6}
\begin{figure}[h!]
\centering
\includegraphics[width=2.2 in]{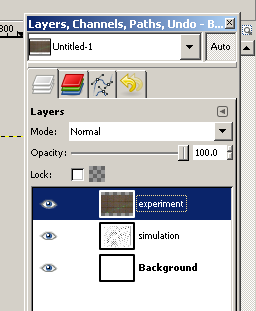}
\caption{A screenshot showing the layer window after adding the `experiment' layer in step \ref{step6}.}
\label{layerwindow}
\end{figure}

\item The layer `experiment' you will notice needs to be scaled. This is where the grids on the screen will help us. Select the layer `experiment' and click on the scale icon from the toolbox window (see Fig. \ref{toolboxscale}).
\begin{figure}[h!]
\centering
\includegraphics[width=2.2 in]{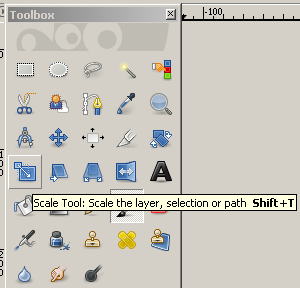}
\caption{A screenshot showing the toolbox window with the \textit{Scale} option.}
\label{toolboxscale}
\end{figure}
\pagebreak

\item The roughly scaled layer `experiment' is shown in Fig. \ref{experimentscale}. The layers window had to be taken out of view for scaling.
\begin{figure}[h!]
\centering
\includegraphics[width=7.0 in]{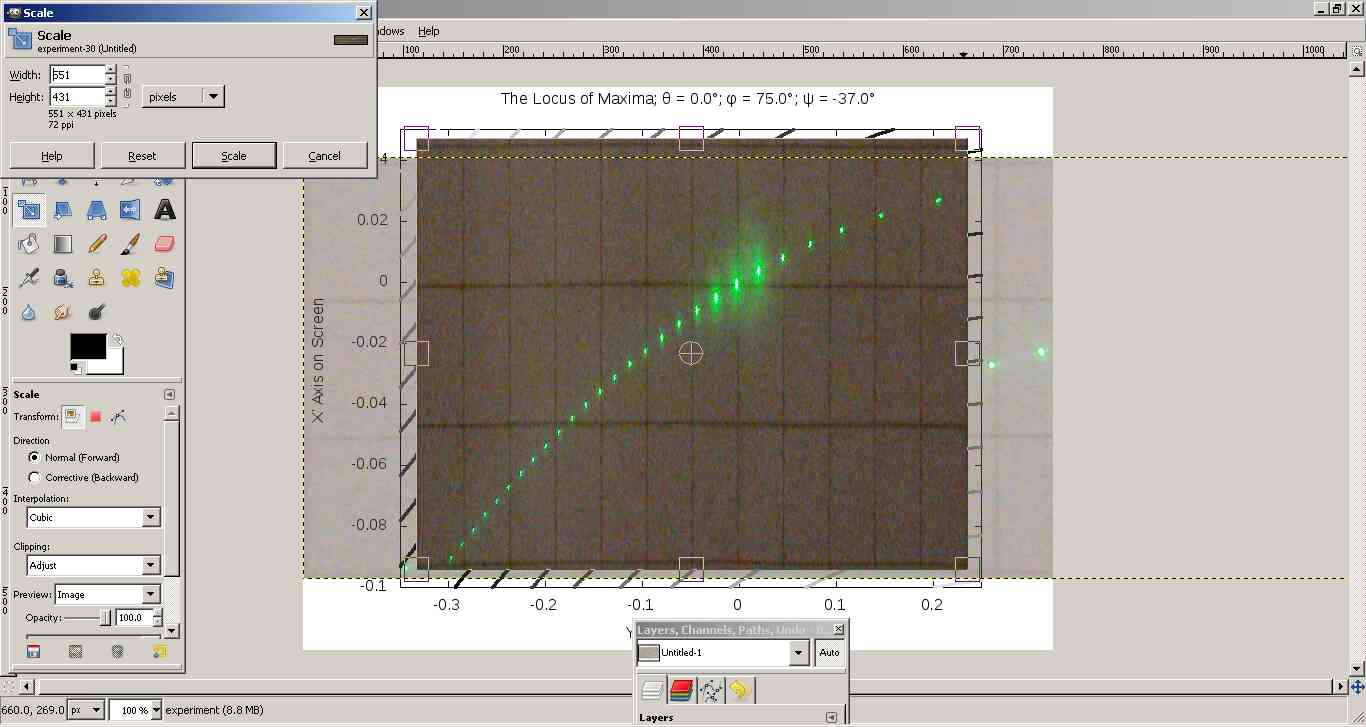}
\caption{A screenshot showing the roughly scaled `experiment' layer.}
\label{experimentscale}
\end{figure}

\item After some fine adjustment for scaling, the scaled overlay with altered opacity appears as shown in Fig. \ref{scaledoverlay}
\begin{figure}[h!]
\centering
\includegraphics[width=7.0 in]{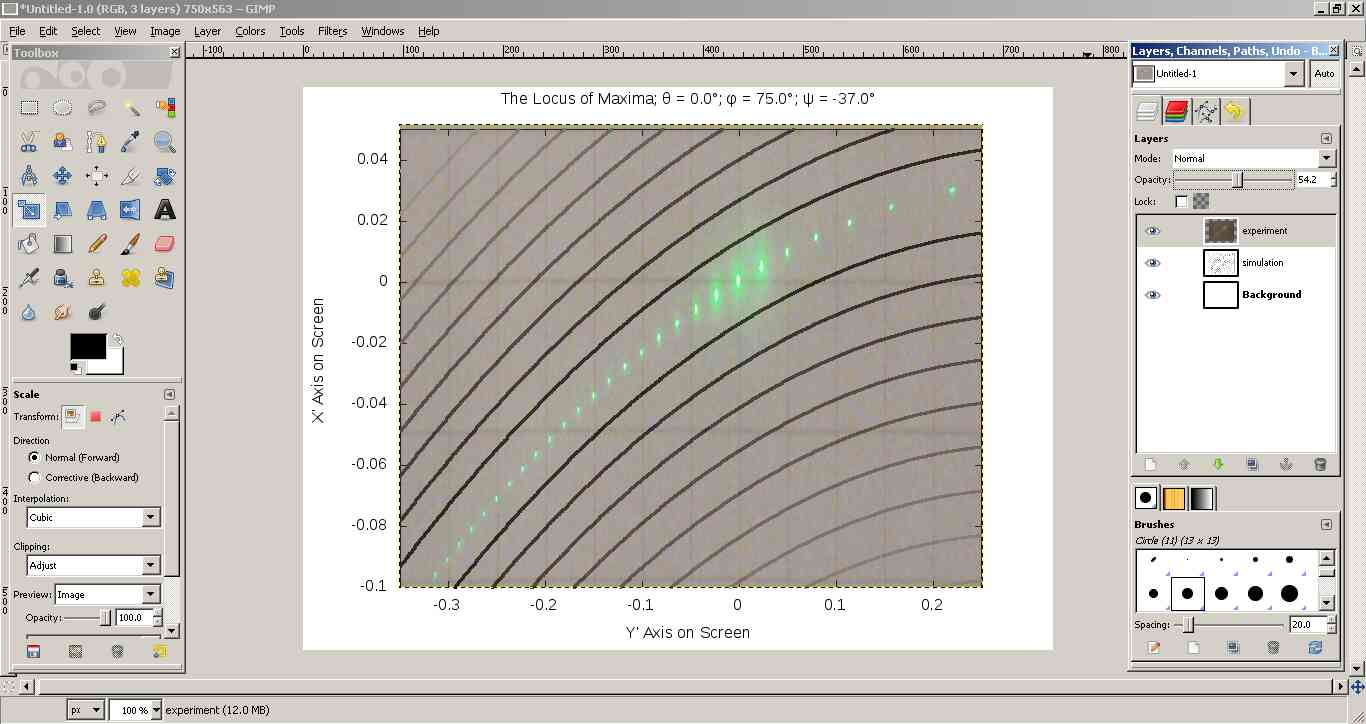}
\caption{A screenshot showing the final scale of the `experiment' layer. Observe the grid coinciding with the axis-tics on the simulated plot.}
\label{scaledoverlay}
\end{figure}
\pagebreak

\item It is still not in the form that seen in the article, we will now edit the colors. First, desaturate (Go to \textit{Colors$\rightarrow$Desaturate}) and select the option \textit{Average}, as shown in Fig. \ref{desaturate}
\begin{figure}[h!]
\centering
\includegraphics[width=7.0 in]{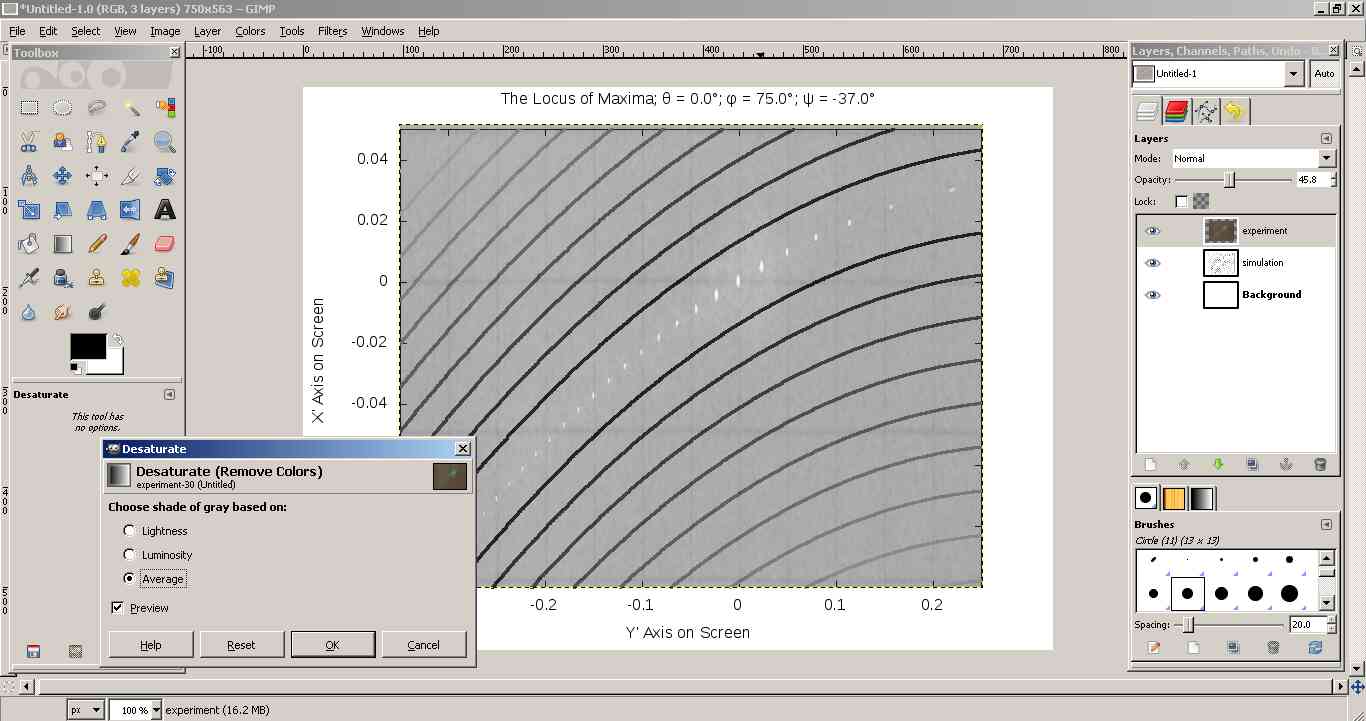}
\caption{A screenshot showing the colors being desaturated in the `experiment' layer.}
\label{desaturate}
\end{figure}

\item Next, invert the colors (\textit{Colors$\rightarrow$Invert}) and then the option `threshold' (\textit{Colors$\rightarrow$Threshold}; adjust the slider to a value between 140 to 150). Finally change the opacity of the experiment layer to about 60, and see the final result of the sample image (Fig. \ref{finalsample}).
\begin{figure}[h!]
\centering
\includegraphics[width=7.0 in]{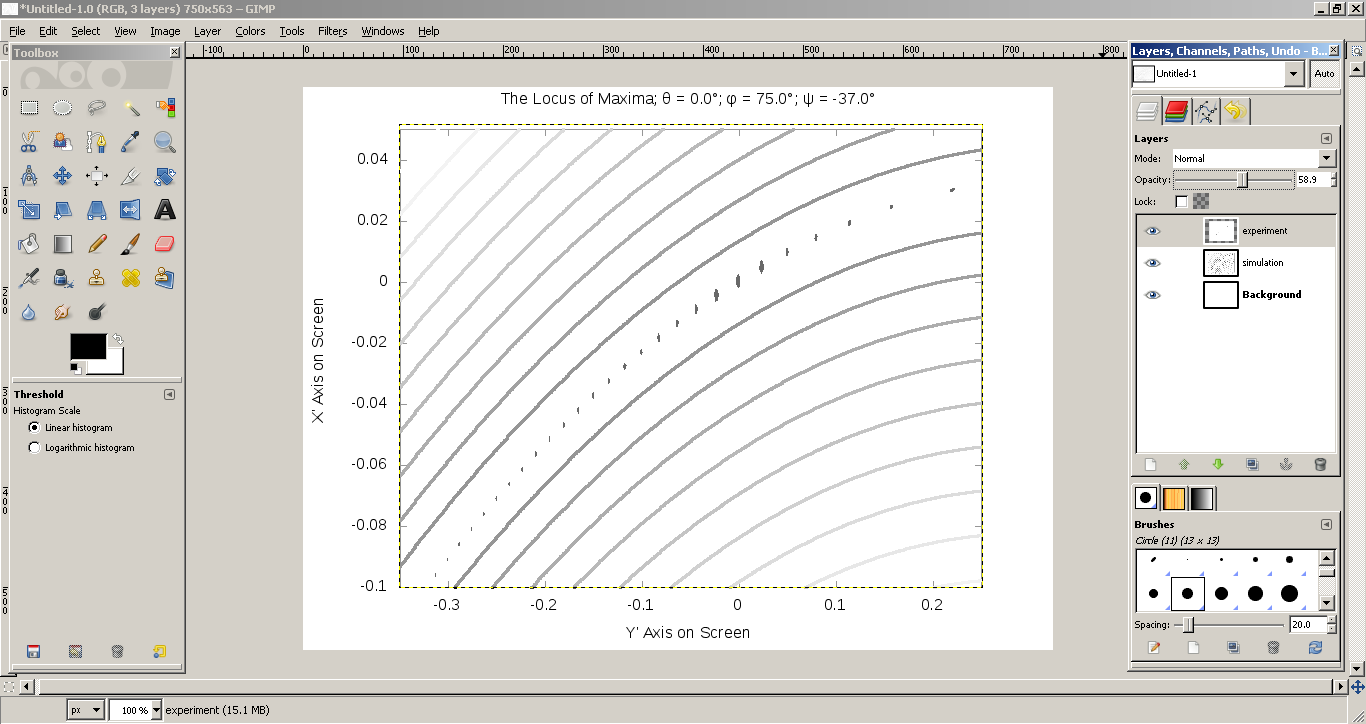}
\caption{A screenshot showing the final state of the sample image after using Invert and Threshold options.}
\label{finalsample}
\end{figure}

\end{enumerate}